\documentclass[12pt]{article}
\pdfoutput=1

\usepackage{cite}
\usepackage{booktabs}
\usepackage[english]{babel}
\usepackage{amsmath,amssymb,amsbsy,amstext, amsthm, simplewick}
\usepackage[backref=page]{hyperref} 
\usepackage{graphicx}
\usepackage{amsfonts}
\usepackage{amssymb}
\usepackage[small]{caption}
\usepackage{upgreek}
\usepackage[svgnames,dvipsnames,x11names,table]{xcolor}
\usepackage{multirow}
\usepackage{geometry}
\usepackage[hang,flushmargin]{footmisc}
\usepackage{bm}
\usepackage{braket}
\usepackage{subcaption}
\usepackage{mathtools}
\usepackage{setspace}
\usepackage{cleveref}
\usepackage{comment}
\usepackage{scalerel}
\usepackage[normalem]{ulem}
\usepackage{slashed}
\usepackage{enumitem}
\usepackage{dsfont}
\usepackage{tikz}
\usetikzlibrary{decorations.markings}
\usetikzlibrary{shapes.misc}
\usetikzlibrary{arrows.meta}

\makeatletter
\g@addto@macro\bfseries{\boldmath}
\makeatother

\hypersetup{
    colorlinks=true,
    linkcolor={red!50!black},
    citecolor={blue!50!black},
    urlcolor={blue!80!black}
}

\usepackage{colortbl}

\makeatletter
\newlength{\apb@width}
\newcommand{\autoparbox}[2][c]{\settowidth{\apb@width}{#2}\parbox[#1]{\apb@width}{#2}}

\makeatother

\definecolor{lightgray}{gray}{0.9}

\usepackage[framemethod=default]{mdframed}
\newmdenv[skipabove=7pt,
skipbelow=7pt,
rightline=false,
leftline=false,
topline=false,
bottomline=false,
backgroundcolor=gray!10,
linecolor=gray,
innerleftmargin=5pt,
innerrightmargin=5pt,
innertopmargin=5pt,
innerbottommargin=5pt,
leftmargin=0cm,
rightmargin=0cm,
linewidth=4pt]{eBox}

\usepackage[most]{tcolorbox}
\tcbset{colback=white, colframe=black,
        highlight math style= {enhanced, 
            colframe=red,colback=red!10!white,boxsep=0pt}
        }
\definecolor{light-gray}{gray}{0.95}

\crefname{table}{Table}{Tables}
\crefname{equation}{Eq.}{Eqs.}
\crefname{appendix}{App.}{Apps.}
\crefname{section}{Sec.}{Secs.}
\crefname{figure}{Fig.}{Figs.}

\numberwithin{equation}{section}

\def\beq{\begin{equation}}
\def\eeq{\end{equation}}

\def\bea{\begin{eqnarray}}
\def\eea{\end{eqnarray}}

\def\d{{\rm d}}

\def\Tr{{\rm Tr}}

\def\beq{\begin{equation}}
\def\eeq{\end{equation}}
\def\bea{\begin{eqnarray}}
\def\eea{\end{eqnarray}}

\def\d{{\rm d}}

\def\Tr{{\rm Tr}}

\def\O{{\cal O}}

\def\Mpl{M_{\rm pl}}

\def\d{{\rm d}}

\def\k{{\vec{\scaleto{k}{7pt}}}}

\def\x{{\vec x}}

\def \nn {\nonumber}

\DeclareRobustCommand{\SkipTocEntry}[4]{}

\definecolor{colorTC}{rgb}{.2,.7,.2}

\definecolor{amethyst}{rgb}{0.6, 0.4, 0.8}

\definecolor{acolor}{rgb}{0.4, 0.2, 0.4}

\definecolor{blue3}{RGB}{31, 119, 180}
\definecolor{red3}{RGB}{	214, 39, 40}
\definecolor{orange3}{RGB}{255, 127, 14}
\definecolor{green3}{RGB}{44, 160, 44}

\begin{document}

\begin{titlepage}
\setcounter{page}{1} \baselineskip=15.5pt
\thispagestyle{empty}
$\quad$
\vskip 70 pt

\begin{center}
{\LARGE \selectfont \bf The Quantum Mechanics of Rare Events \\[4pt]}
{\large \selectfont \bf From Quantum Walks to Stochastic Inflation \\[4pt]}
\end{center}

\vskip 20pt
\begin{center}
\noindent
{\fontsize{12}{18}\selectfont Daniel Green, Kshitij Gupta, and Akhil Premkumar }
\end{center}

\begin{center}
\vskip 4pt
\textit{{\small Department of Physics, University of California at San Diego,  La Jolla, CA 92093, USA}}

\end{center}

\vspace{0.4cm}
 \begin{center}{\bf Abstract}
 \end{center}

\noindent 

Rare fluctuations in physical systems depend on the detailed microphysics responsible for the fluctuations. In classical statistical systems, the large deviation principle has elucidated the role of semi-classics in describing this regime, and has simultaneously provided a the mathematical foundation of statistical mechanics. Large deviation theory for quantum system is considerably less developed. As all physical systems are fundamentally quantum mechanical, this leaves a major gap in our understanding of rare fluctuations relevant to statistical physics, cosmology, and more. In this paper, we develop the practical aspects of the theory of large deviations relevant for calculating rare events in physical systems from quantum walks to cosmology. We first analyze the case of the anharmonic oscillator coupled to a bath, showing explicitly how the system evolves from dominantly statistical (e.g. thermal) to quantum fluctuations. We then generalize these results, showing that the dominant rare fluctuations minimize the measurement-induced relative entropy. This perspective provides a thermodynamic description of a wide range of open quantum systems. We apply these results to random walks that arise in cosmology through stochastic inflation. We show that the evolution of the density matrix of long wavelength fields on a fixed de Sitter background breaks the KMS symmetry, giving rise to a stationary density matrix that does not respect detailed balance.



\end{titlepage}
\setcounter{page}{2}

\restoregeometry

\begin{spacing}{1.2}
\newpage
\setcounter{tocdepth}{2}
\tableofcontents
\end{spacing}

\setstretch{1.1}
\newpage

\section{Introduction}

Physics predicts the outcomes of measurements in terms of statistical averages. In classical physics, the source of randomness, or noise, is chaotic microphysics that is not resolved.  In quantum mechanics, in addition to any unresolved microphysics, randomness is also a fundamental ingredient through Born's rule. Yet, when calculating statistical averages of physical observables, the techniques used in quantum mechanics and classical statistical mechanics are similar~\cite{Feynman:1963fq,Nelson:1966sp,Parisi:1982ud,Caldeira:1982iu}. Typically, the predictions of a theory are expressed in terms of correlation functions of a small number of operators, which ultimately encode the shape of the probability distribution around its peak. 

The calculation of rare events, or the tail of the distribution, is of a fundamentally different character than low-order correlation functions. In complex systems, there are usually only a small number of configurations with a high probability of occurring, but there are many ways of arriving in a highly unlikely configuration. The probability of an exponentially unlikely event is typically determined by the least unlikely way to arrive in that unlikely situation. 

Rare events are also difficult to understand using numeric techniques alone. For typical fluctuations in statistical systems, it is often easy (in principle) to extract the properties of the distribution using stochastic simulations. However, simulating a single rare fluctuation typically requires exponential resources, making accumulating sufficient statistics impractical. Instead, semi-classical methods offer a concrete path to making analytic predictions in this regime, as they are naturally exponentially small. However, many semi-classical methods involve truncations of the full theory, making it unclear if they are accounting for the relevant physics~\cite{Coleman:1977py,Hartle:1983ai,Hawking:1981fz,Hong:2003pe,2016JSP...162..793B,Brown:2017cca}.

Despite these challenges, understanding both tails of distributions and semi-classics is both observationally relevant and theoretically compelling. Observationally, structures from galaxies~\cite{Press:1973iz,Bardeen:1985tr} to primordial black holes~\cite{Carr:1974nx,Carr:2025kdk} are formed from large but rare density fluctuations. In addition, the productions of large numbers of particles, or high point correlation functions, are ultimately tied to non-trivial saddles~\cite{Libanov:1994ug,Son:1995wz}. Theoretically, semi-classical configurations have proved valuable in understanding the structure of theories at large charge~\cite{Hellerman:2015nra,Monin:2016jmo} and/or large operator dimension~\cite{Gubser:2002tv}. More dramatically, the interpretation of semi-classical gravity remains tied to some of the most basic questions in quantum gravity~\cite{Almheiri:2020cfm,Harlow:2022qsq}.

For classical stochastic systems, large deviation theory has provided a rigorous mathematical framework for understanding the behavior of rare fluctuations for a wide range of phenomena~\cite{Varadhan_2008,Touchette:2009mis,2025arXiv250316015B}. The intuition is most easily understood in the context of classical random walks, particularly those where the steps are independent and identically distributed (i.i.d.). Given such a walk with $N$ random steps, the probability of finding the walker at a distance $x = N \alpha$, where $\alpha ={\cal O}(1)$, obeys the large deviation principle when it is given by
\beq
P(x= N \alpha) \propto e^{-N I(\alpha)}  \ ,
\eeq
where the rate function, $I(\alpha)$, is independent of $N$. The scaling with $N$ is ubiquitous in physical examples but is non-trivial nonetheless. For example, a single step of a Gaussian random walk covering a distance $N$ would scale as $e^{-{\cal O}(N^2)}$. In this sense, rare events are typically a consequence of \textit{a collective behavior of $N$ steps}, rather than being sensitive to any single step~\cite{Cohen:2022clv}.

The collective behavior that gives rise to this universal $N$-scaling, also means that the rate functions themselves are determined by ``instantons": the most probable way to arrive at a rare event is due to fluctuations around a smooth path where all the steps are correlated. Mathematically, the Gärtner-Ellis theorem~\cite{doi:10.1137/1122003,ellis1984large} provides a rigorous demonstration of this intuition, as the rate functions are determined by minimization of the effective action defined by the generating function. In practice, these are often solutions to a classical set of equations of motion, and have enabled the calculation of rare events in a wide range of physical systems~\cite{2005JSP...119..677B,Grafke:2013ska,2015JPhA...48G3001G,2017JPhA...50z3001A,2018PNAS..115..855D,2018PNAS..115...24R,2019JSP...179.1637R,Woillez:2019xwc,2019PhRvX...9d1057D,2021NCimR..44..291G,2026arXiv260612624D}.

The implications of the large deviation principle are more profound when expressed in terms of the measured distribution of steps, rather than just the total distance. Specifically, for an $N$-step walk, suppose we ask for the probability of finding $n_s$ occurrences of a step size $s$ that should occur with a probability $\nu_s$. Sanov's theorem~\cite{sanov1957large} tells us the probability of finding a given distribution of steps, $\{ n_s \}$, is given in terms of the relative entropy between distributions $n_s/N \to\mu(s)$ and $\nu_s \to \nu(s)$, $S_{\rm rel}(\mu||\nu)$ or, equivalently, the Kullback-Leibler (KL) divergence $D_{\mathrm{KL}}(\mu \| \nu)$,
\beq
S_{\rm rel}(\mu \| \nu) =  D_{\mathrm{KL}}(\mu \| p)=\sum_{s \in \mathcal{S}} \mu(s) \log \frac{\mu(s)}{\nu(s)} \ ,
\eeq
where $\mathcal{S}$ is the set of possible step sizes.
This connection can be seen by direct calculation, where summing over walks with the same distribution of steps in different permutations yields the probability for a given distribution,
\beq
P(\{n_s \}) = \exp\left(- N \sum_{s \in \mathcal{S}} \frac{n_s}{N} \log \frac{n_s}{N \nu_s} \right) \to \exp\left(- N S_{\rm rel}(\mu \| \nu)\right)\ .
\eeq
Not only do we see the appearance of the relative entropy, minimizing $S_{\rm rel}$, while imposing the constraint $\sum_{s \in \mathcal{S}} n_s s = x$, reproduces the rate function for $x = N\alpha$, $I(\alpha)$, as calculated by the Gärtner-Ellis theorem\footnote{We will review this argument in Section~\ref{sec:entropy}.}. In this precise sense, large deviations in a classical random walk minimize the relative entropy.

\begin{figure}[!ht]
    \centering
    \includegraphics[width=0.6\textwidth]{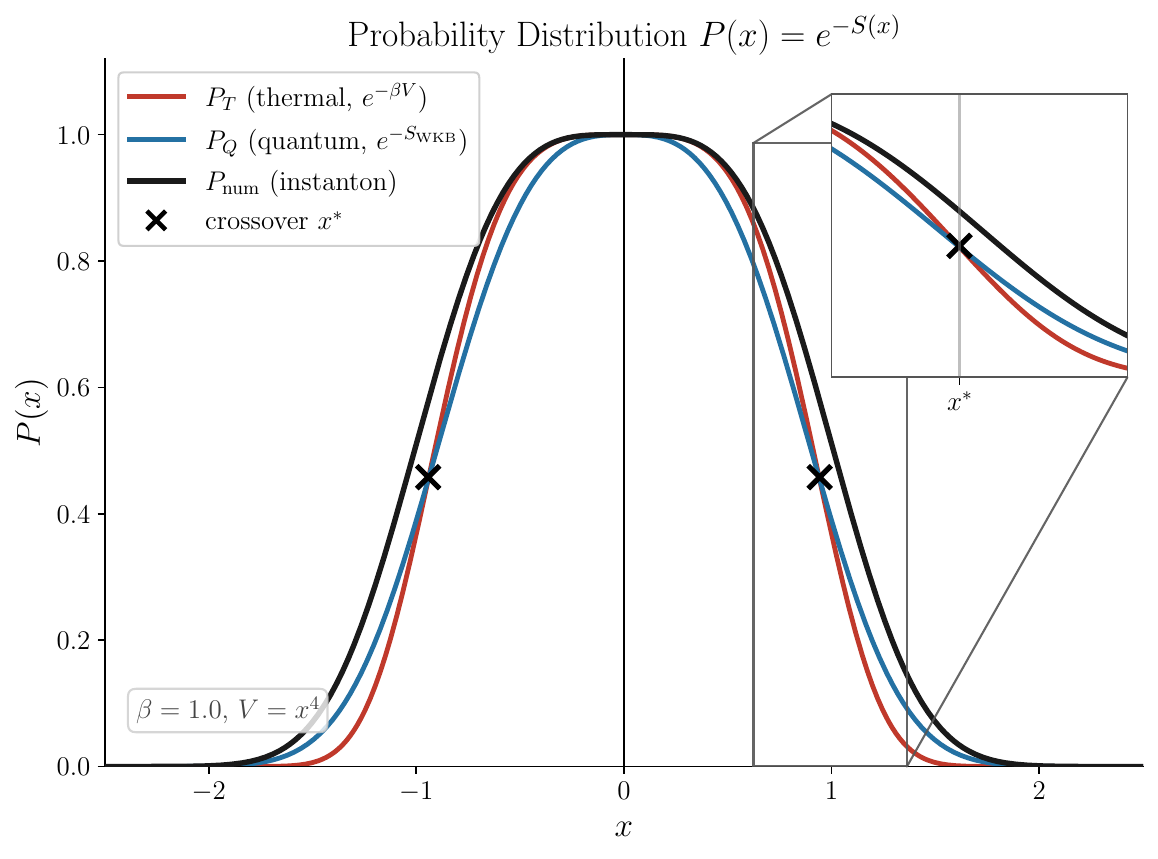}
    \caption{Illustration of the relative contributions of classical random (thermal, red) and quantum mechanical (WKB, blue) contributions to the full probability distribution (numerical instanton, black) at large distances calculated. The calculation for all three curves are explained in Section~\ref{sec:qw}.
    }
    \label{fig:summary}
\end{figure}

For quantum mechanical systems, the same depth of understanding is still lacking (but see e.g.~\cite{Hayashi:2002dfe,Hayashi:2012rfj,Znidaric:2014vzm,Znidaric:2014rmm,Gherardini:2015tde,Bringuier:2017dia,Carollo:2018fei,Gherardini:2019plm,Denzler:2020bga,Paulino:2024egn,Miangolarra:2025byc,Albert:2026tvd,Cao:2026vgu} for recent developments). We might have expected a quantum generalization of Sanov's theorem to express probabilities for quantum walks in terms of the quantum relative entropy. Several such theorems exist with applications that include quantum state estimation~\cite{Keyl:2005hkv} and open quantum systems~\cite{Bjelakovic:2004wqp,vanHorssen:2014dqi,Carollo:2021ega,Hayashi:2024wlf,Lami:2025ybb}. However, it is not straightforward to apply these theorems to the practical application of calculating the probability of rare events. For a quantum mechanical generalization of a random walk, or quantum walk~\cite{Venegas-Andraca:2012zkr}, we cannot simply replace the number $n_s/N$ with a density matrix. Care is required to determine the correct physical quantity. This is representative of the general challenge defining large deviations in quantum systems: we might have hoped to replace every instance of a probability with a density matrix, but this fails when we have to account for specific measured outcomes like number counts. These issues become particularly relevant when discussing the tails of the distribution, where competition between quantum and statistical effects become important (see \cref{fig:summary}).

Our goal in this paper is to advance our understanding of the tails of distributions in quantum statistical systems for practical calculations, particularly with an eye towards quantum field theory.  We will focus specifically on quantum walks, as these are the simplest examples, and yet they have a wide range of applications physical problems of interest. We will first consider the problem of a quantum anharmonic oscillator at finite temperature. Close to the minimum of the potential, the equilibrium distribution is given by the thermal behavior, while at larger distances, it is given by the WKB approximation of the wavefunction. We will derive these behaviors, and the transition from one behavior to the other, using multiple techniques to confirm the intuition. These results are closely related to previous works on tunneling in finite temperature systems~\cite{Noble:1979yf,Affleck:1980ac,Athron:2023xlk}. We will then generalize the results to non-thermal models of the noise, which have not been  explored in the literature but appear when describing quantum field theory in de Sitter space~\cite{Nambu:1991vs,Li:2025azq,Green:2025hmo,Calderon-Figueroa:2025dto,Cespedes:2026fdp,Li:2026lwl,Christie:2026dwx}.

Based on the experience with the anharmonic oscillator, we will demonstrate that the transition from thermal to quantum behavior can be understood as resulting from maximizing the quantum fidelity, at the level of individual paths or, relatedly, minimizing measurement-induced relative entropy for the distribution. We show specifically that the measurement-induced relative entropy between the empirical density matrix (defined by the distribution of steps) and true density matrix plays the role of number counts in Sanov's theorem for this particular process. From this point of view, the transition from quantum to classical tails is given purely by this minimization procedure. The quantum relative entropy itself does not appear because coupling from the walker to the quantum system picks a preferred basis. This allows the walker to generate entropy even when the steps are defined by a pure state. The quantum behavior of the walker does not always dominate the behavior of the tail, but does in many examples where the tail of the stochastic fluctuations is known to be more suppressed, or equivalently, that the statistical fluctuations have a parametrically smaller fidelity, or larger measurement-induced relative entropy.

Although the main results apply to general quantum system, the choice of examples is inspired by problems in cosmology where the local physics is well approximated by a random (quantum) walk. This description, known as stochastic inflation~\cite{Vilenkin:1983xq,Starobinsky:1986fx}, is a powerful tool for understanding the behavior of light fields in accelerating cosmologies. The fields locally acquire a highly non-Gaussian probability distribution that is well understood using the Fokker-Planck equation~\cite{Starobinsky:1994bd}. However, for rare fluctuations, this framework fails as the probability becomes sensitive to the microphysics of the random fluctuations~\cite{Cohen:2021jbo,Cohen:2022clv}. More importantly, the quantum nature of this random walk becomes important in exactly the same regime as the failure of the perturbative expansion of the classical noise. After developing both the anharmonic oscillator and entropic interpretation of general rate functions for quantum systems, we return to the implications for cosmology. This perspective clarifies the meaning of corrections to stochastic inflation and its relationship to quantum field theoretic calculations in pure de Sitter space. Moreover, it points to the physical original of the failure of the Kubo–Martin–Schwinger~\cite{Kubo:1957mj,Martin:1959jp,Haag:1967sg} (KMS) condition for the density matrix, also recently observed in out-of-time-ordered correlators (OTOC)~\cite{Milekhin:2026tbi,Cui:2026bcd,Chen:2026boh,Harlow:2026pwe}.

This paper is organized as follows: In Section~\ref{sec:qw}, we introduce quantum walks and apply the large deviation principle to the anharmonic oscillator. In Section~\ref{sec:entropy}, we discuss quantum walks on a more abstract level and use the relationship between entropy and large deviations to understand the behavior of a general quantum walk. We apply these ideas to cosmology and stochastic inflation in Section~\ref{sec:cosmology}, and conclude in Section~\ref{sec:conclusions}.

\begin{table}[htbp]
\centering
\footnotesize
\renewcommand{\arraystretch}{1.15}
\setlength{\tabcolsep}{4pt}
\begin{tabular}{@{}l p{0.30\linewidth} @{\hspace{1.4em}} l p{0.30\linewidth}@{}}
\toprule
\multicolumn{2}{l}{\textbf{Random walks }} &
\multicolumn{2}{l}{\textbf{Open quantum systems}} \\
\midrule
$N$ & Number of steps in the walk
 & $\rho(x,x')$ & Density matrix of the walker \\
$s_i$ & Size of the $i$-th step
 & $\rho_0$ & Initial density matrix \\
$x,\,x_0$ & Final / initial position of the walker
 & $\chi(s_i,s_i')$ & Density matrix associated with the $i$th step \\
$\nu(s)$ & True (single-step) distribution of step sizes
 & $x_c,\,x_q$ & Classical / quantum (Keldysh) variables \\
$\mu(s)$ & Empirical distribution of step sizes
 & $F(\Sigma\Vert\chi_N)$ & Quantum fidelity \\
$q(x_0)$ & Distribution of the initial position
 & $D_{{\cal M}}(\sigma\Vert\chi)$ & Measurement-induced relative entropy in basis ${\cal M}$ \\
$\alpha$ & Rescaled position, $x=N\alpha$
 & $\Sigma,\,\sigma$ & Measurement projector / empirical density matrix \\
$I(\alpha)$ & Large deviation rate function & $S_{\rm e}, S_{\rm hk}$ & Excess / housekeeping entropy \\
$D_{\mathrm{KL}}(\mu\Vert\nu),\,S_{\mathrm{rel}}(\mu\Vert\nu)$ & Kullback--Leibler / relative entropy & $s_{\rm tot}, S_{\rm tot}$& Total entropy for a path / distribution \\
\bottomrule
\end{tabular}
\caption{Summary of notation used throughout the paper.}
\label{tab:symbols}
\end{table}

\newpage

\section{Quantum Walks and the Anharmonic Oscillator}\label{sec:qw}

In this section, we will study large fluctuations in simple quantum walks. One of the main goals will be to understand the tail of the probability distribution in the position basis for the anharmonic oscillator, 
\beq
H = \frac{p^2}{2m} + \frac{1}{2} m \omega^2 x^2 + \frac{1}{4}\lambda x^4 \ ,
\eeq
both at finite temperature, and in the presence of classical noise / dissipation. The general construction is similar to previous work on quantum Brownian motion~\cite{Caldeira:1981rx,Grabert:1988yt,Hu:1991di}, but with a specific focus on large fluctuations. This model is known to exhibit non-trivial but distinct tails both classically at finite temperature and quantum mechanically from the wavefunction. As a result, the scaling behavior of the tail as a function of $x$ will precisely reveal the relative importance of classical and quantum fluctuations. This model will therefore serve as valuable intuition for a more general understanding of tails in open quantum systems, including scalar field theory in de Sitter space~\cite{Nambu:1991vs,Li:2025azq,Green:2025hmo,Calderon-Figueroa:2025dto,Cespedes:2026fdp,Li:2026lwl,Christie:2026dwx}. The examples in this section are essentially identical to the behavior of a local particle detector in a cosmological setting, up to higher order corrections discussed in Section~\ref{sec:cosmology}.

\subsection{Basic Quantum Walks}

We are interested in how rare fluctuations arise in quantum systems with many degrees of freedom. Like in the case of classical stochastic systems, to understand the general framework it is helpful to have the example of simple random walks in mind as a reference. We will therefore introduce a class of quantum random walks that are minimal generalizations of classical walks to provide this kind of intuition.

The most basic example is to consider a walker as a free particle with a wavefunction in the position basis, $\psi_0(x)$. For simplicity, we will assume for now that $\psi_0(x-c)$ is a valid time-independent wavefunction for all values of $c$. Now we couple our walker, $x$, to a collection of $N$ particles whose wavefunctions, $\tilde \psi(s_i)$, in their position basis $s_i$, are identical so that the wavefunction of the whole system is
\beq\label{eq:quantum_walk}
|\Psi \rangle = e^{i \hat p \sum_i s_i} \psi_0(x) \otimes \prod_{i=1}^N\tilde \psi(s_i) \ .
\eeq
We see here that the particles $s_i$ play the role of the steps in a random walk that translate the wavefunction $\psi_0(x)$ of the walker. Because the original wavefunction $\psi_0(x)$ is stationary even after translations, the position of the walker remains in a stationary state that is shifted due to the entanglement with the walkers, $\tilde \psi(s_i)$.

There is no reason this procedure needs to be limited to pure states. We can apply the same intuition in terms of an initial density matrix for our walker, $\psi_0(x) \psi^*(x') \to \rho_0(x,x')$, and density matrices of the steps, $\tilde \psi(s_i) \tilde \psi^*(s_i')\to \chi(s_i,s_i')$ so that the reduced density matrix for our walker is 
\beq\label{eq:qwdensity}
\rho(x,x') = \bigg\langle x \bigg| {\rm Tr}_i \exp \left(i \sum_i \hat p \hat s_i \right) \rho_0 \otimes \chi^{\otimes N}  \exp \left(-i \sum_i \hat p \hat s_i \right) \bigg| x' \bigg\rangle \ .
\eeq
We will take the case where the `noise', i.e., the $N$ particles $s_i$ the walker is coupled to, is Gaussian. We now define our quantum walk the following way: at each instant in time, we couple to a new $\chi$, and trace it out. This is the quantum equivalent of a classical walker undergoing $\dot{x} = \xi$ where $\xi$ is a Gaussian stochastic noise. In this case, the density matrix evolves according to \cite{Green:2025hmo} 
\begin{equation}\label{eq:qheat}
    \frac{\partial \rho}{\partial t} = g\left(\frac{\partial}{\partial x} + \frac{\partial}{\partial x'}\right)^2 \rho(x,x') \ ,
\end{equation}
which is the quantum equivalent of the diffusion equation. This can be seen most directly when the density matrix is diagonal, $\rho(x,x') = P(x) \delta(x-x')$, where the combination of derivatives is needed for \cref{eq:qheat} to reduce to the heat equation for $P(x)$ with no additional terms from derivatives of $\delta$-functions. The relationship between this quantum walk and an i.i.d.~random walk is derived explicitly in Appendix~\ref{app:iid}.

We can further understand the nature of the evolution equation when the density matrix in the Keldysh variables, $x_c = (x + x')/2, x_q = x - x'$, so that
\begin{equation}
    \frac{\partial \rho}{\partial t} = g \frac{\partial^2 }{\partial x_c^2} \rho \left(x_c + \tfrac{1}{2} x_q , x_c - \tfrac{1}{2} x_q \right) \ .
\end{equation}
In these variables, the evolution of the density matrix is simply the heat equation in $x_c$ and time-independent in $x_q$. Here $x_c$ and $x_q$ denote the classical and quantum behavior in the $x$-basis, as $x_c$ and $x_q$ determine the diagonal and off-diagonal components of $\rho$ respectively. The heat equation being purely in terms of $x_c$ is therefore consistent with our expectations for classical noise diffusion.

Another common representation for understanding this equation is given via transforming to the Wigner function. The Wigner function is a quasi-probability distribution, defined by Fourier transforming the off-diagonal elements of the density matrix
\begin{equation}
    W(x_c,p_q) = \int d x_q \, e^{-i p_q x_q} \rho(x_c, x_q)
\end{equation}
In terms of the Wigner distribution, the heat equation is given by 
\begin{equation}
    \frac{\partial W}{\partial t} = g \frac{\partial^2 W}{\partial x_c^2}
\end{equation}
Importantly, $W(x,p)$ is not necessarily positive throughout it's domain, and hence can't be though of as a true probability distribution. On marginalizing one of the variables, for example $p$, we do get the true probability distribution along $x$, via $P(x) = \int dp \, W(x,p)$ etc. While we stick to working with density matrices for now, we will come back to the Wigner descriptions in later sections. 

Now, one can add further add a Hamiltonian dynamics to the walker so that we can allow for unitary time evolution and response to the steps. The density matrix then evolves according to
\begin{equation}
\begin{aligned}
    \frac{\partial \rho}{\partial t} &= -i[H, \rho] + \frac{g}{2} \frac{\partial^2 \rho}{\partial x_c^2} \\
    &= i \frac{\partial^2 \rho}{\partial x_c \partial x_q}  - i [V(x_c + \tfrac{1}{2} x_q) - V(x_c - \tfrac{1}{2} x_q)] \rho + \frac{g}{2} \frac{\partial^2 \rho}{\partial x_c^2}  \ .
\end{aligned}
\end{equation}
Note that this structure is just the open system Lindbladian, defined via
\begin{equation}
    \frac{\partial \rho}{\partial t} = -i[H, \rho] + \sum L^\dagger \rho L + \frac{1}{2} \{L^\dagger L, \rho \} \ .
\end{equation}
In the case of our diffusion model, we have $L = i \sqrt{g}\hat{P}$.

It will be useful to also have a path integral representation of the same density matrix, known as the open influence functional (see e.g.~\cite{Sieberer:2015svu,Haehl:2016pec,Hongo:2018ant,Liu:2018kfw,Akyuz:2023lsm,Colas:2025app,Kaplanek:2026kpp,Pajer:2026fuo} for discussion). For a closed system, the path integral, under a general Hamiltonian $H = p^2/2m + V(x)$ is given by
\begin{equation}
\begin{aligned}
    \rho_{\rm closed} &= \int {\cal D} x {\cal D}x' \exp(i S[x] - iS[x']) \rho(t_0) \\
    &= \int  {\cal D} x {\cal D}x' \exp \left( i\int dt\, \left[ \frac{1}{2}m (\dot{x}^2 - \dot{x}'{}^2) + V(x) - V(x') \right] \right) \rho(t_0) \ .
\end{aligned}
\end{equation}
This is, of course, just two copies of the path integral representation of the ground state (under suitable $i\epsilon$ prescriptions for the time integrals). Next, we would like to back to pure diffusion and introduce it to our path integral for the density matrix. For simplicity, we will take the case of diffusion in momentum, achieved via the operator $e^{i \hat x s_i}$. We see that on integrating out one particle $s_i$, whose distribution is $\chi(s_i, s_i) \propto \exp(-s_i^2/2g)$, we get 
\begin{equation}
    \rho(x,x') = \int dq\, e^{i (x - x') q} e^{-q^2/2g}\rho_{\rm closed}= Ae^{-g(x - x')^2/2} \rho_{\rm closed}(x,x') \ ,
\end{equation}
where $\hat x \rho = x\rho$ and $x'\rho = \rho \hat x$, are the superoperators acting from the left and the right. To represent a continuous diffusion, we repeat this process at each time step so that
\begin{equation}
\begin{aligned}
    \rho &= \int {\cal D}x {\cal D}x'\int {\cal D}s\, \exp \left(i S_{\rm closed} +i \int dt\, (x(t) - x'(t)) s(t) \right) \exp\left(-\int dt\, \frac{s^2(t)}{2g} \right) \rho_0\\
    &={\cal N}\int {\cal D} x {\cal D}x' \exp \left(i S_{\rm closed}-g\int dt\, \frac{(x(t) - x'(t))^2}{2} \right) \rho_0 \ , 
\end{aligned}
\end{equation}
where $S_{\rm closed} \equiv S[x] - iS[x']$.
As a result, adding the diffusion at each time step due to these extra states, $\chi$, gives us\footnote{We reabsorb the normalization constant ${\cal N}$ into the path integral measure throughout the paper. Alternatively, the normalization is fixed by $\Tr \rho = 1$.} 
\begin{equation}
\begin{aligned}
    \rho &= \int  {\cal D} x{\cal D}x' \exp\left( i\int dt\, \left[ \frac{1}{2} m (\dot{x}^2 - \dot{x}'{}^2) + V(x) - V(x') - g\frac{(x - x')^2}{2}\right] \right)\rho(t_0) \ .
\end{aligned}
\end{equation}
Going to the Schwinger-Keldysh variables, we get
\begin{equation}
     \rho = \int  {\cal D} x {\cal D}x' \exp\left( i\int dt\, [ m \dot{x}_c \dot{x}_q + V(x_c + \tfrac{1}{2} x_q) - V(x_c - \tfrac{1}{2} x_q)] - \int dt\, \frac{gx_q^2}{2} \right)\rho(t_0) \ .
\end{equation}
This is a simple example of the open influence functional, which describes the quantum version of the stochastic equation
\begin{equation}
    \dot{x}_c = \frac{p_q}{m} \qquad \dot{p}_q = -V'(x_c) + \sqrt{g} \xi \ .
\end{equation}
where $\xi$ is white noise with variance $\langle \xi(t) \xi(t') \rangle = \delta(t-t')$. We use $x_c, p_q$ for the quantum version of the classical Langevin equation to emphasize that the resulting phase space evolution still obeys quantum mechanical rules of non-commutation, measurement-induced collapse etc, which will be important in our later discussion. In the next few sections, we will use this open influence functional method to describe various quantum walks and understand their rare events. 

\subsection{Finite Temperature from Random Walks}

One can use the open influence function to model a particle undergoing stochastic fluctuations in a bath. Classically, a particle in a potential $V(x)$, when subject to  random noise and friction, equilibrates to the Gibbs distribution. The equations of motion in this case is given by
\begin{equation}
    \dot{x} = \frac{p}{m} \qquad \dot{p} = -\gamma p - V'(x) + \xi_T \ .
\end{equation}
where $\xi_T$ is a Gaussian random noise, with variance 
\begin{equation}
    \langle \xi_T(t) \xi_T(t')\rangle = 2\gamma m k_B T \delta(t - t') \ ,
\end{equation}
such that the fluctuation-dissipation theorem is satisfied. Turning to the quantum description, the density matrix can be represented by the path integral~\cite{Sieberer:2015svu,Haehl:2016pec,Hongo:2018ant,Liu:2018kfw,Akyuz:2023lsm,Colas:2025app,Kaplanek:2026kpp,Pajer:2026fuo}, namely
\begin{equation}\label{eq:Schwinger_Keldysh_thermal}
    \begin{aligned}
    iS &= i \int dt\, [ m \dot{x}_{c} \dot{x}_{q} - m\gamma \dot{x}_{c} x_{q} - (V(x_{c} +  \tfrac{1}{2}x_{q}) - V(x_c - \tfrac{1}{2} x_q) ] -  \frac{m\gamma}{\beta} \int dt\, x_q^2 \ .
    \end{aligned}
\end{equation}
The two non-unitary terms are the $m \gamma \dot{x}_c x_q$ term, which represents damping due to friction and the $x_q^2$ term which represents the random force. Since our primary interest is the probability distribution in the $x$ basis, $P(x) =\rho(x,x)$, we are interested in the diagonal terms in the density matrix which in these variables means the boundary conditions at the final time, $t_f$, are $x_c(t_f)= x, x_q(t_f) = 0$.

To use the techniques of large deviations, we proceed as follows. Our goal is to find a semi-classical solution that respects our boundary conditions for the density matrix. If $x_q(t) \ll x_c(t)$ along the entire semi-classical trajectory, we can expand out the $(V(x_{c} +  \frac{1}{2}x_{q}) - V(x_c - \frac{1}{2} x_q))$ terms in the action as
\begin{equation} \label{eq:action_classical_thermal}
    \begin{aligned}
    iS &= i \int dt\, [ m \dot{x}_{c} \dot{x}_{q} - m\gamma \dot{x}_{c} x_{q} - x_q V'(x_c) ] -  \frac{m\gamma}{\beta} \int dt\, x_q^2 \ .
    \end{aligned}
\end{equation}
This is the well known classical MSR action \cite{MSR} for a particle in a heat bath. In this description, we can integrate out $x_q$ since it is purely Gaussian. The corresponding result for $x_c$ is the same as a completely classical random walk due to thermal fluctuations. As is well known, such a walk has the equilibrium distribution defined by $P(x) \propto \exp(-\beta V(x))$. 

Given our goal of understanding the interplay of quantum and classical effects, we should understand how to derive the equilibrium distribution, $\exp(-\beta V(x))$, as a saddle point of \eqref{eq:action_classical_thermal} when including both $x_c$ and $x_q$. These saddles should correspond to solutions of the equations of motion derived from the path integral, 
\begin{equation}
    m\ddot{x}_{c} + m\gamma \dot{x}_{c} + V'(x_{c}) = 2i m\frac{\gamma }{\beta}  x_q \qquad 
    m\ddot{x}_{q} - m\gamma \dot{x}_{q} + V''(x_{c})x_{q} = 0 \ . \label{eq:ThermalWalkEOM}
\end{equation}
There are two types of solutions to these equations. The first is where we trivially satisfy both the equations of motion and the boundary conditions for $x_q$ using $x_q(t) = 0$. In this case, we have the set of solutions for $x_c$ from the equation 
\begin{equation}
     m\ddot{x}_{c} + m\gamma \dot{x}_{c} + V'(x_{c}) = 0 \qquad x_q =0  \ .
\end{equation}
For $x_c$, this describes the particle experiencing a damping friction and a potential $V(x)$. Since we are interested in explaining instantons that start from the bottom of the potential $V(x)$, these paths cannot describe the motion we are interested in. 

To find the other solution, we can use the KMS symmetry of the thermal instantons.  One can explicitly check that the action and hence the equations of motion are unchanged under the transformation $\tilde x_c(t) = x_c(-t), \tilde x_q(t)= x_q(-t) - i\beta \dot{x}_c(-t)$, meaning the set of solutions for $\tilde x_{c,q}$ and $x_{c,q}$ are identical. This symmetry is a consequence of thermality (periodicity in Euclidean time) and time translation invariance. It also enforces detailed balance, and that time-forward and reversed paths have equal probability.

For our purposes, KMS is important because the individual solutions are not invariant under the transformation, even though the equations of motion are.  As a result, KMS transforms our first class of solutions into a second class. Specifically, the solution $x_q(t) = 0, x_c(t)$ implies the existence of a second solution where $\tilde x_q(t) = - i \beta \dot{x}_c(t)$.
Again, one can also simply plug this in as an ansatz and confirm that solving the resulting equations of motion for $x_c$, 
\begin{equation}\label{eq:thermal_instanton_influence}
    m \ddot{x}_c - m\gamma \dot{x}_c + V'(x_c) = 0 \qquad x_q = -i\beta \dot{x}_c \ ,
\end{equation}
will produce a valid solution to the original equations of motion.

Crucially, the second set of solutions describes antidamping for $x_c$. This is the time reversed classical equation, and describes fluctuations which can climb up the potential. As we will now show, this solution reproduces the Gibbs distribution when plugged into the action \eqref{eq:action_classical_thermal}. Substituting $x_q$ back in, and using the $x_c$ equations of motion, we get
\begin{equation}\label{eq:thermal_eq}
\begin{aligned}
    iS &= -\beta \int dt\, [m \dot{x}_c \ddot{x}_c + m\gamma \dot{x}_c^2 +V'(x_c) \dot{x}_c] + \frac{m\gamma}{\beta} \int dt\, \beta^2 \dot{x}_c^2 \\ 
    &= -\beta m\gamma \int dt\, \dot{x}_c^2 = -\beta \int dt\, \dot{x}_c (\ddot{x}_c + V'(x_c))\\
    &= -\beta \int dt\, \frac{1}{2}\frac{d}{dt}(\dot{x}_c^2) - \beta \int dt\, \frac{d}{dt} (V(x_c))\\
    &= -[\beta V(x_f) - \beta V(x_0)] \ ,
\end{aligned}
\end{equation}
where $x_0$ and $x_f$ are the initial and final positions, and we take the initial and final velocities to be zero. Thus, putting the minima of the potential at $V(x_0) = 0$, we get the Gibbs result, $P(x) \propto \exp(-\beta V(x))$. 

This construction helps us understand the validity of assuming that the walk is classical. In going from the Schwinger-Keldysh action \eqref{eq:Schwinger_Keldysh_thermal} to the classical action \eqref{eq:action_classical_thermal}, we assumed that $x_q \ll x_c$. On solving the EOM, we got $x_q = - i\beta \dot{x}_c$. Thus, we require
\begin{equation}
    \beta \dot{x}_c \ll x_c \implies \beta \omega_x \ll 1 \ ,
\end{equation}
where $\dot{x}_c = \omega_x x_c$ is the time scale associated with $x_c$ at some point. For a general potential, it is given by the 
\begin{equation}
    \omega_x = \sqrt{\frac{V''(x_c)}{m}} \ .
\end{equation}
Thus, when $\beta \omega_x \sim 1$, we can no longer assume that $x_q \ll x_c$, and hence the off-diagonal terms of the density matrix play a role. Thus the regime of validity of classical description of the walk is 
\begin{equation}\label{eq:estimateopen}
    \beta \sqrt{\frac{V''(x_c)}{m}} \ll 1  \ ,
\end{equation}
beyond which the tails become quantum. In the next section, we use the thermal circle to derive a similar inequality.

\subsection{Finite Temperature from the Thermal Circle}

The most straightforward approach to understanding the anharmonic oscillator at finite temperature is to use the Euclidean action for a quantum particle on $S^1$ with our thermal (periodic) boundary conditions imposed on (Euclidean) time. Since we are interested in the regime of exponentially small probabilities, we can just calculate the density matrix using semi-classical solutions (thermal instanton) to the action
\beq
\rho(x,x') = \exp \left[ - \int_{-\beta/2}^{\beta/2} dt_E \left( \frac{1}{2} m \dot x{}^2 + V(x) \right) \right] \ ,
\eeq
where we impose $x(0_+) =x$ and $x(0_-) = x'$, where here $\dot x$ is now derivative with respect to $t_E$. Using the equations of motion, as usual, we can find
\beq
m \ddot x =   V' \to \frac{m}{2} \dot x{}^2 =  V(x) - \epsilon \ ,
\eeq
where $-\epsilon$ is the energy. The instanton contribution to the density matrix for a given $\epsilon$ then becomes 
\beq
\rho(x,x')  = \exp\left[ -\beta \epsilon -\int_x^{x'} d\tilde x \sqrt{2m(V(\tilde x)- \epsilon)} \right] \ .
\eeq
Since $\epsilon$ is just a constant of integration, we should find the value that maximizes the probability, which yields the constraint
\beq
\beta =  \int_x^{x'} d\tilde x \frac{\sqrt{2m}}{\sqrt{V(\tilde x)- \epsilon}} \ .
\eeq
In principle, this is a closed set of equations that can be solved for any $x$ and $x'$ (which is, of course, the same as solving the equations of motion).

Given the formal solution, we would like to evaluate these integrals to confirm that the density matrix is consistent with our thermal expectations. First, we will focus on the probability distribution where $x=x'$ where the boundary conditions simplify. In order to make sense of this regime, one defines the turning point $x=a$ so that $\epsilon = V(a)$. Now we have
\beq
- \log P(x) = \beta V(a) - \int_a^x d\tilde x \sqrt{2m(V(\tilde x)- V(a))} , \qquad \beta =  \int_x^{a} d\tilde x \frac{\sqrt{2m}}{\sqrt{(V(\tilde x)- V(a))}} \ .
\eeq
In this situation, the high temperature limit is clear. When $\beta \to 0$, we need to take $a = x$ and therefore $P(x)\to e^{-\beta V(x)}$ as we would expect from high temperature. In contrast, taking $\beta \to \infty$ ($T \to 0$) is a bit more obscure. If we assume\footnote{To have a self-consistent result with $\beta \to \infty$, we need $V(\tilde x)-V(a)$ to vanish {\it at least} quadratically in $(x-a) \to 0$. This requires that the slope vanishes at $\tilde x = a$, but places no additional restrictions on the shape of the potential.} that near the minimum of the potential $V(x)-V_{\rm min} \approx (x-x_{\rm min})^2$ then the integral that determines $\beta$ will be logarithmically divergent, so that
\beq\label{eq:WKB}
- \log P(x) = \beta V_{\rm min} - \int_{x_{\rm min}}^x d\tilde x \sqrt{2m(V(\tilde x)- V_{\rm min})} \ .
\eeq
which is the quantum result in the ground state. Notice that away from the extrema, the integral converges if the slope near $x=a$ is non-zero.

The transition between these two regimes can be estimated from the value of $x=x_*$, where $a=x_*$ and $a=x_{\rm min}$ give qualitatively similar results. To simplify, let's set $V(x_{\rm min}=0$, so that the cross-over occurs when
\beq
\beta V(x_*) \approx \int^{x_\star}_{x_{\rm min}} d\tilde x \sqrt{2m V(\tilde x)} \approx x_\star \sqrt{2m V(x_\star)} \to  \beta \sqrt{\frac{V(x_*)}{2m x_*^2} } \approx 1 \ .
\eeq
As a qualitative estimate, we see that the overall scaling is the same as Equation~(\ref{eq:estimateopen}) since, at the level of our above estimate, $V'' \approx V/x^2$.

\begin{figure}[!ht]
    \centering
    \includegraphics[width=0.95\textwidth]{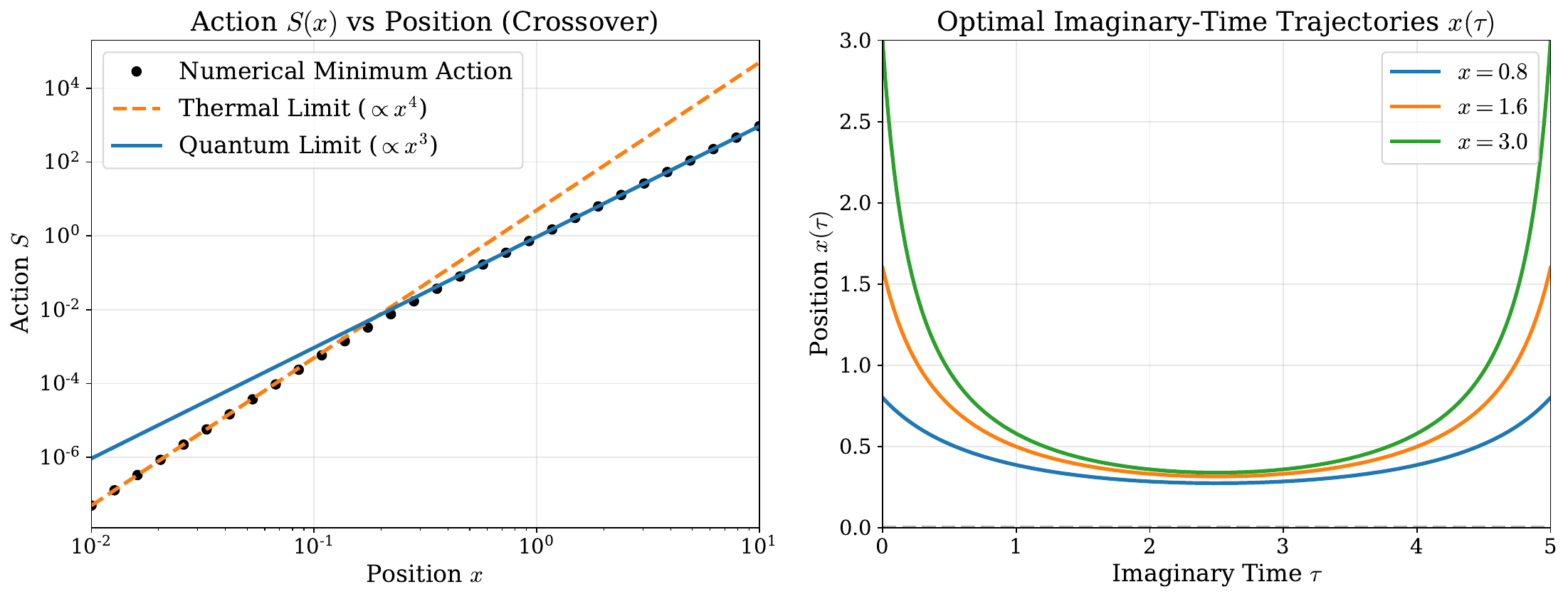}
    \caption{{\it Left:} Numerically computed action for the quantum thermal instanton for various values of $x$ (points), compared to the thermal and quantum scalings, $S \propto x^4$ and $x^3$ respectively. {\it Right:} The solutions for the instantons for different boundary conditions.
    }
    \label{fig:instanton}
\end{figure}

We can numerically evaluate the action for any $x$ to confirm this cross-over behavior. In Figure~\ref{fig:instanton}, we compare these numerical solutions for $V(x) = x^4$ with $m=1$ and $\beta =5$.

Now, the claim is that for the density matrix, we have two possible saddles. First, we could just integrate from the same bounce $x$ to $x'$, 
\beq
\rho_1(x,x') =\exp \left( -\beta \epsilon -\int_x^{x'} dx \sqrt{2m(V(x)- \epsilon)} \right) \ ,
\eeq
The other one involves splitting the integral into paths from $a$ to $x$ and from $a$ to $x'$,
\bea\label{eq:WKB_rho}
\rho_2(x,x')  &=& \exp \left( -\beta \epsilon -\int_a^x d\tilde x \sqrt{2m(V(\tilde x)- \epsilon)}-\int_a^{x'} d\tilde x \sqrt{2m(V(\tilde x)- \epsilon)} \right) \ .
\eea
This would be just the product of the two results we found before for the WKB wavefunction, which is fitting for a state that is more quantum than classical.

Given these two type of saddle, we understand the transition for the thermal to the quantum behavior at large $x$ as being related to the transition between these saddles. For sufficiently large $x$, we can lower the action by having the path approach $x = a$, even though the path itself is longer. When this occurs, the density matrix no long looks diagonal and we recover pure state description from the WKB wavefunction.

\subsection{General Random Walks}\label{sec:noise_in_x}

Having understood the thermal case well, we now turn our attention to the tails of more general walks. We will study the saddle points of $x_c$ and $x_q$ in an open influence action to understand the regime in which $x_q \ll x_c$. By doing so, we can discern the scale at which tails transition from classical to quantum behavior.

We consider a random walk where we have friction in $p$ and noise in $x$. Such a walk is seen in cosmological settings \cite{Green:2025hmo}. First, we need to derive the quantum influence functional for such a model. The Langevin equation is given by
\begin{equation}\label{eq:CosmoToyEOM}
    \dot{x}_c = \frac{p_q}{m} + \sqrt{g} \xi \ , \qquad  \dot{p}_q = -\gamma p_q - V'(x_c) \ , 
\end{equation}
which when reduced to one variable, is given by
\begin{equation}
 \begin{aligned} 
\ddot{x}_c + \gamma \dot{x}_c + \frac{V'(x_c)}{m} = \gamma \sqrt{g} \xi + \sqrt{g} \dot{\xi} \ .
 \end{aligned} 
\end{equation}
which suggests \footnote{Note that mathematically, the time derivative of white noise $\dot{\xi}$ is not a well defined quantity. The equation is just meant to give intuition for the final  action.} that the influence functional will be given by
\begin{equation}\label{eq:action_noise_in_x}
\begin{aligned}
   Z &= \int \mathcal{D}x \exp\bigg( i \int dt\, \left[ m \dot{x}_{q} \dot{x}_{c} - m\gamma \dot{x}_{c} x_{q} - \left(V(x_c + \tfrac{1}{2} x_q) - V(x_c - \tfrac{1}{2} x_q ) \right) \right]\\
   & \qquad \qquad - \int dt\, \frac{ gm^2\gamma ^{2} }{2}x_{q}^{2} - \int dt\, g m^2 \frac{\dot{x}_{q}^{2}}{2}  \bigg) \ . \\
\end{aligned}
\end{equation}
We give a rigorous derivation in Appendix \ref{app:action_derivation}. Since we first want to understand the classical tails, we can expand out the potential terms via $x_q \ll x_c$, giving 
\begin{equation}
    Z = \int \mathcal{D}x \exp\bigg( i \int dt\, \left[ m \dot{x}_{q} \dot{x}_{c} - m\gamma \dot{x}_{c} x_{q} - V'(x_c)x_q \right] - \int dt\, \frac{g m^2 }{2}\left(\gamma ^{2} x_{q}^{2} + \dot x_q^2\right)  \bigg) \ .
\end{equation}
The saddle points are given by 
\begin{equation}
 \begin{aligned} 
m\ddot{x}_{c} + m\gamma \dot{x}_{c} + V'(x_{c}) &= i gm^2\gamma ^{2} x_{q} - igm^2 \ddot{x}_{q} \ ,\\
m\ddot{x}_{q} - m\gamma \dot{x}_{q} + V''(x_{c}) x_{q} &= 0 \ .
 \end{aligned} 
\end{equation}
While there does not exist a closed form solution to this equation like for the Gibbs case, we can still extract key features analytically under certain regimes. 

First, we take the strongly overdamped regime, where we can neglect the inertia of $x_c$ generically. This approximation is valid as long as $V''(x_c)/m \gamma^2 \ll 1$. In the thermal case, we saw that the path that minimizes the action is given by the time reversal of the classical path. In particular, we can neglect the inertia in the overdamped limit for both $x_c$ and $x_q$. Hence, we can drop the $\ddot{x}_q$ term relative to the $\gamma^2 x_q$ term, so that
\begin{equation}
\begin{aligned}
    m\ddot{x}_{c} + m\gamma \dot{x}_{c} + V'(x_{c}) &= i gm^2\gamma ^{2}x_{q} \ , \\
m\ddot{x}_{q} - m\gamma \dot{x}_{q} + V''(x_{c}) x_{q} &= 0 \ .
\end{aligned}
\end{equation}
This is the same equations of motion as the thermal distribution case, \eqref{eq:ThermalWalkEOM}, and we can write down the solution as 
\begin{equation}
 \begin{aligned} 
x_{q} = -\frac{2i}{\gamma m g} \dot{x}_{c} \ .
 \end{aligned} 
\end{equation}
Comparing with the thermal solution $x_q = -i \beta \dot{x}_c$, we see that the effective temperature is $\beta = 2/\gamma m g$. The condition for validity of the overdamped thermal tails becomes 
\begin{equation}
    \frac{1}{m\gamma g}\dot{x}_{c} \ll x_{c} \implies \frac{V'(x_{c})}{m^2\gamma^2 g} \ll x_{c} \implies \frac{V''(x_{c})}{m^2 \gamma ^{2} g} \ll 1 \ .
\end{equation}
We started with the assumption that we are in the overdamped regime, which requires $\frac{V''(x_c)}{m \gamma^2} \ll 1$, and hence we need that 
\begin{equation}
    m g \gtrsim  1 \ .
\end{equation}
Generically, we will satisfy this condition if we assume $g$ and $\gamma$ both depend on some common quantity (for instance in de Sitter, $ g, \gamma \propto H^3, H$), so that $g$ is large when $\gamma$ is large. Thus, we have shown that for $V''(x_c)/m \gamma^2 \ll 1$, we satisfy $x_q \ll x_c$ and can therefore treat the tails as classical.

\begin{figure}[htbp]
    \centering
    \includegraphics[width=0.6\textwidth]{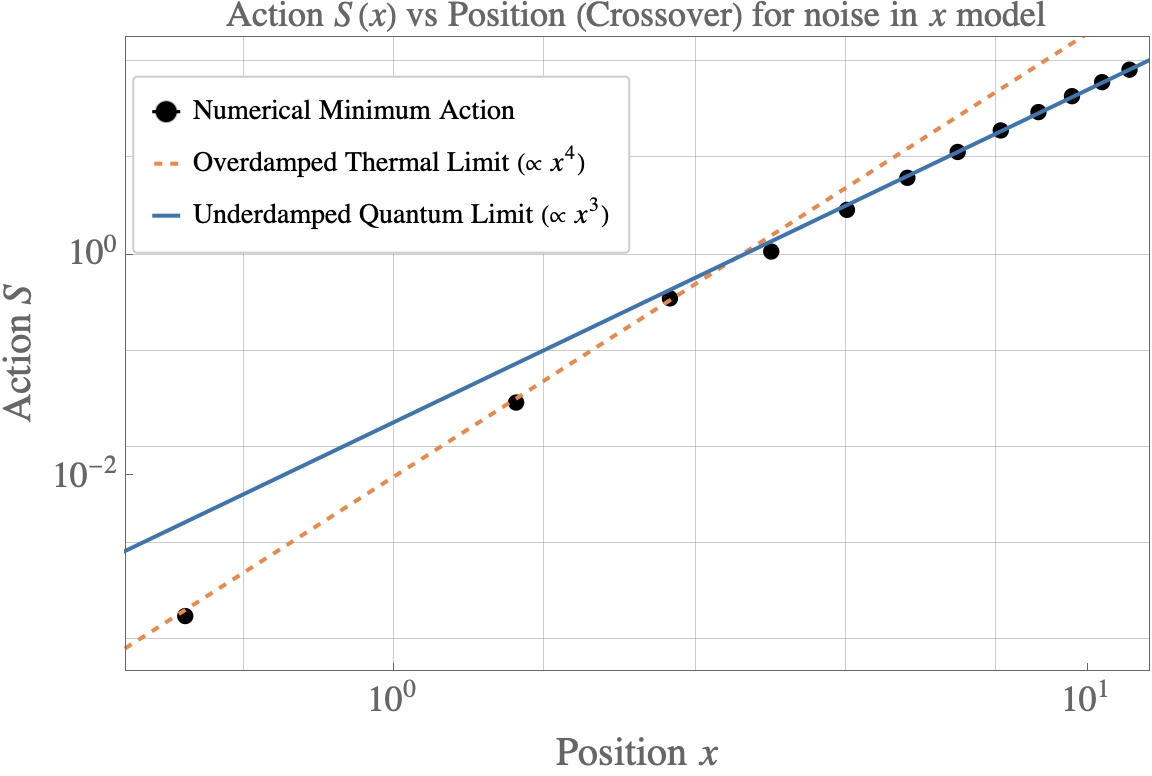}
    \caption{Numerically computed action for the quantum noise in $x$ instanton for various values of $x$ (points), compared to the overdamped classical and free quantum scalings, $S \propto x^4$ and $x^3$ respectively. We took $V(x) = \lambda x^4/4$, with $m = 1 = \hbar, \gamma = 5, \lambda = 2/5, g = 4$.}
    \label{fig:instanton_noise_in_x}
\end{figure}

Second, at some large value of $x_c$, $V''(x_c) \gg m \gamma^2$ and our overdamped limit is no longer valid and we will find the result is dominated by the under-damped regime. Then, assuming a common origin for $g$ and $\gamma$, we can neglect the $\sqrt{g} \xi$ and $-\gamma p$ terms in \eqref{eq:CosmoToyEOM} which returns us to a pure state description. Thus, we see that interpolating between overdamped and underdamped limit gives us transitions between classical and quantum tails. Numerical verification of this cross-over is shown in Figure~\ref{fig:instanton_noise_in_x}.

More generally, we see that large deviations of general quantum walks can be determined using straightforward semi-classical techniques. While closed-form analytic solutions are difficult to find for these more general walks, the is no general obstacle to numeric calculations. In principle, if $\chi$ and $\rho_0$ are completely specified, we can always find the tail of the distribution this way. However, when the wavefunction is only known perturbatively, like the examples described in Section~\ref{sec:cosmology}, at sufficiently large $x,x'$ the perturbative description can break down.

\section{Relative Entropy and Quantum Large Deviations}\label{sec:entropy}

Quantum mechanics presents a non-trivial change to the behavior of random walks because it introduces a second source of randomness that is unrelated to the classical noise. The probability of rare events is therefore a competition between these two sources of randomness and allows for transitions as one or the can dominate in different regimes. We saw this explicitly in the case of the anharmonic oscillator. Naturally, we would like to understand the interplay between these two effects at a deeper level. In classical random walks, Sanov's theorem shows that the probability of rare fluctuations arises from optimal paths that minimize the relative entropy. Our goal in this section is to generalize this understanding to quantum walks.

\subsection{A Classical Model}
\label{sec:ClassicalModel}

The challenge of competing random effects is not limited to quantum walks, but can be easily modeled in classical systems. For concreteness, suppose we have an i.i.d.~random walk comprised of steps $\{ s_i \}_{i=1}^{N}$, each drawn from a distribution $\nu(s)$. In addition, we also assume that the initial position $x_0$ is drawn from a distribution $q(x_0)$. Intuitively, the probability of finding yourself at $x$, defined by 
\beq\label{eq:class_model}
x = x_0 + \sum_i^N s_i \qquad x_0\in q(x_0) \quad s_i \in \nu(s)\ ,
\eeq
is the probability of $N$ steps moving a distance $x-x_0$ times the probability of start at $x_0$. In equations, this means
\beq
P(x, x_0 |N) = q(x_0) P(x-x_0|N) \qquad P(x|N)  = \left( \prod_i^N  \int ds_i \, \nu(s_i) \right) \delta \left( \sum_{i=1}^{N} s_i - x \right) \ .
\eeq
Clearly the probability of a rare event will involve competition between the tail of $q(x_0)$ leading to a large initial displacement, versus the tail of $P(x|N)$ allowing the walk to cover a large distance over the $N$ steps. Our goal is to connect this to the minimum entropy path, so we will take a detour through Sanov's theorem.

The are many ways the walker can traverse a distance $x$ in $N$ steps. For instance, the walker could make one large large jump $s_i \approx x$ at some specific instant $i$, with all other steps $s_i$ averaging to nearly 0, \textit{or} take concerted steps $s_i \approx x/N$ for all $i$. Therefore, we pose the following question: \textit{given} that the walker has covered a distance $x$ in $N$ steps, what is the most probable distribution of steps it took to get there?\footnote{The precise order of the steps is immaterial, so we focus on the distribution of step sizes.}

Given a probability distribution $\nu(s)$, the probability that we observe an {\it empirical} distribution of steps $n_s /N \to\mu(s)$, where $n_s$ is the number count of steps of size $s$, is given by
\beq\label{eq:sanov}
P[\mu(s)] =  \exp \left( -N \int ds \, \mu(s) \log \frac{\mu(s)}{\nu(s)} \right) \ .
\eeq
This formula is more intuitive in the case of discrete steps, where
\beq
P(\{n_s\}) \propto N! \prod_{s\in {\cal S}} \frac{\nu^{n_s}_s}{n_s!} \to \exp\left(- N \sum_{s\in {\cal S}} \frac{n_s}{N}\log\frac{n_s}{N \nu_s}\right) \ .
\eeq 
The same result holds in the continuum limit $n_s/N \to \mu(s)$.
The integral is the KL divergence $D_{\rm KL}(\mu \| \nu)$, which measures the ``distance'' between those distributions; clearly any $\mu(s)$ that is different from $\nu(s)$ is exponentially suppressed. By imposing the constraint that $N$ steps from $\mu(s)$ cover a distance $x$, we can compute the probability $P[\mu(s)|x,N]$ over that specific subset of $\mu$'s. Since there is an exponential penalty for deviating from $\nu(s)$, it follows that $P[\mu(s)|x,N]$ is sharply peaked at a $\mu_\star(s)$ that is as ``close'' as possible to $\nu(s)$ while also satisfying the constraint. In the large $N$ limit, most of the probability mass in $P[\mu(s)|x,N]$ concentrates in a very narrow region around $\mu_\star(s)$, therefore $\mu_\star(s)$ is very nearly the \textit{only} way of reaching $x$ in $N$ steps.\footnote{A similar argument explains the emergence of macroscopic determinism from microscopic chaos in thermodynamics \cite{Jaynes65}. The same idea was extended to nonequilibrium settings by Schr\"odinger \cite{Chetrite21}.} In other words, $P(x|N) = P[\mu_\star(s)]$.

Operationally, we maximize the probability from \eqref{eq:sanov} subject to the constraints that $\mu(s)$ is normalized and $\sum s_i = x \equiv N \alpha$. Equivalently, we can minimize $-\log P[\mu(s)]$ under the same constraints,
\beq
- \log P(x|N)= {\min}_{\mu,\lambda, \nu} \left[\int ds \left( \mu(s) \log \frac{\mu(s)}{\nu(s)} + \zeta (s \mu(s) -\alpha) + \xi (\mu(s) -1) \right) \right] \ , \label{eq:Likelihood}
\eeq
where $\zeta$ and $\xi$ are Lagrange multipliers. Differentiating with respect to $\mu(y)$ gives
\beq
\log \frac{\mu(y)}{p(y)} + 1 - \zeta y -\xi = 0 \to \mu(y) = p(y) e^{\zeta y +\xi -1} \ .
\eeq
We now impose the constraints from minimizing with respect to the Lagrange multipliers to find
\beq
e^{\xi -1} \int ds \nu(s)e^{\zeta s} = 1 \to e^{1-\xi} =  \int ds \nu(s)e^{\zeta s}  \equiv f(\zeta) \ , \label{eq:NormalizationConstraint}
\eeq
and 
\beq
e^{\xi -1} \int ds s \nu(s)e^{\zeta s}  = \frac{\partial}{\partial \zeta} e^{\xi -1} \int ds \nu(s)e^{\zeta s} = \frac{f'(\zeta) }{f(\zeta)} = \alpha \ . \label{eq:MeanConstraint}
\eeq
If we define $\Lambda = \log f$, then $\Lambda'(\zeta) = \alpha$. Solving these equations, we can determine $\zeta_\star$, $\nu_\star$, and $\Lambda_\star$ for a given $\alpha$. Plugging these into \eqref{eq:Likelihood} leads directly to Cram\'er's theorem~\cite{zbMATH02516150},
\beq
-\log P(x|N) = \int ds \mu_\star(s) \log \frac{\mu_\star(s)}{\nu(s)} = \int ds \nu(s) e^{\zeta_\star s - \Lambda_\star} (\zeta_\star s -\Lambda_\star) = \zeta_\star \alpha - \Lambda_\star \ ,
\eeq
where $\Lambda = \log \int dx p(x) e^{\zeta x}$ is just the moment generating function for a single step.\footnote{Cram\'er's theorem is essentially the result of Gärtner-Ellis theorem~\cite{doi:10.1137/1122003,ellis1984large} applied to the special case of i.i.d.~random walks. The key difference is that Cram\'er's theorem is expressed in terms of the moment generating function of a single step in the walk, $s_i$, rather that for the entire walk, $x$.} In the large deviation literature~\cite{Varadhan_2008,Touchette:2009mis,2025arXiv250316015B}, the probabilities for number counts ($n_s$) and sample means ($x/N$) are referred to as ``level 2'' and ``level 1'' large deviation principles respectively. The procedure of moving from level 2 to level 1 through minimizing is an example of the contraction principle.

Now that we understand how $P(x|N)$ arises from minimizing the relative entropy, we can introduce $x_0 = N \alpha_0$. We have some rate function for the $N$-step random walk from $x_0=N \alpha_0$ to $x= N \alpha$, call it $I(\alpha-\alpha_0)$, and a probability $q(x_0)$ for starting at $x_0$. So the dominant probability will be the one that maximizes the product of the two probability,
\beq\label{eq:class_prob}
-\log P(x, x_0 | N) = {\rm min}_{x_0} \left(-\log q(x_0) - N I(\alpha-\alpha_0)\right) \ .
\eeq
Notice that $-\log q(x_0)$ is the relative entropy for finding a single entry at $x_0$. In other words, the $q(x_0)$ term is the same as \cref{eq:sanov} but for the case $N=1$ (since we only draw the initial condition once). At the same time, we noticed already that $I(\alpha-\alpha_0)$ is itself the minimum of the relative entropy of the walk. So collectively, the probability of finding the walker at $x$ is given in terms of the minimum of the \textit{total} relative entropy 
\beq\label{eq:Smindef}
-\log P(x, x_0 |N) = {\rm min}_{  \mu, x_0} S_{\rm R} (x_0,  \mu| q(x_0),\nu) \ ,
\eeq
where $\mu(s) = n_s/N$ is the empirical distribution of $s_i$ values and we have defined
\beq
 S_{\rm R} (x_0,  \mu| q(x_0),\nu)  = -\log q(x_0) - N \sum_{s\in{\cal S}} \mu(s)\log \frac{\mu(s)}{\nu(s)} \ .
\eeq
The minimization in \cref{eq:Smindef} is subject to the constraint that $\sum_{s\in S}\mu(s) s = x-x_0$. The consequence is that the tail of the distribution in $x$ is determined by minimizing the relative entropy between $x_0$ and the random steps.

This example is qualitatively similar to our quantum walk problem. If we work in a single diagonal basis, the $q(x) \to |\psi_0(x)|^2$ can encode the quantum uncertainty in the initial position of the walker. The challenge then is defining this result in way that does not require a fixed basis. 

\subsection{Quantum Sanov's Theorem and Tails of Wavefunctions}\label{sec:quantum_sanov}

To understand the full quantum generalization of our classical toy model, consider the quantum walk defined in \cref{eq:qwdensity}, namely
\beq\label{eq:q_model}
\rho(x,x') = \bigg\langle x \bigg| {\rm Tr}_i \exp \left(i \sum_i \hat p \hat s_i \right) \rho_0 \otimes \chi^{\otimes N}  \exp \left(-i \sum_i \hat p \hat s_i \right) \bigg| x' \bigg\rangle \ ,
\eeq
where $\chi(s,s')$ is the density matrix that defines each step, $\hat s_i$, taken by our walker. If the initial density matrix, $\rho_0$ is diagonal, then this would be equivalent to our classical example with $\rho_0$ playing the role of $q(x_0)$. So, in this concrete sense, the question is how to define the problem without any unnecessary restrictions on $\rho_0$ or $\chi$. 

For the random walk steps specifically, the trace is over the density matrices $\chi_i$. Due to the trace and the coupling to $s_i$, only the diagonal $\chi$ terms in the $s_i$ basis contribute to our result. We can therefore treat the walk itself as if it were classical. If we work in the momentum basis of the density matrix $\rho_0(p,p')$, the we see 
\beq
\rho(x,x') = \int \frac{dp dp'}{(2\pi)^2} e^{-i p x+ i p' x'} \rho_0(p,p') \left( {\rm Tr} e^{ip s} \chi(s,s') e^{-i p' s'} \right)^N\ \ .
\eeq
Here $s,s',p$, and $p'$ are all real numbers. The trace over $s$ is defined by taking $s=s'$ and integrating, 
\beq
\rho(x,x') = \int \frac{dp dp'}{(2\pi)^2} e^{-i p x+ i p' x'} \rho_0(p,p') \left(\int ds e^{i(p-p')s} \chi(s,s)  \right)^N\ .
\eeq
As shown in Appendix~\ref{app:iid}, this expression is identical to the probability distribution for purely classical steps if we identify the distributions $\nu(s) =\chi(s,s)$ and take the initial state to be $\rho_0 = |x=0\rangle \langle x=0|$.

However, suppose we are interested in the tail of the distribution where $x = N \alpha$. If we have $\rho_0(x,x)\gg \chi(\alpha,\alpha)^N$, then probability of finding $x=N\alpha$ will be determined in terms of the initial probability. Concretely, let us imagine we were in a pure state initially, $\rho_0 = |\Psi\rangle \langle \Psi|$ with a non-Gaussian tail and $\chi(s,s)$ is a normal distribution with variance $\sigma$, then the resulting density matrix should obey
\beq\label{eq:reduced_rho_w}
\rho(x,x') \approx \langle x | \Psi\rangle \langle \Psi | x' \rangle \qquad x,x' \gg N \sigma \ .
\eeq
In this situation, the tail of the density matrix looks like a pure state because it is determined entirely by the tail of $\rho_0$. The consequence is that deriving the tail from some kind of relative entropy will have to allow pure states as a possible outcome. As explained in Appendix~\ref{app:entropy}, the quantum relative entropy of pure states if infinite, suggesting that this is not likely to be the correct quantity for describing the tail.

In order to make progress, let us first understand how Sanov's theorem works at the level of the density matrix (i.e.\ quantum Sanov's theorem), {\it without} assuming we are performing a quantum walk. We will still assume we have $N$ identical copies of a density matrix $\chi$ for a discrete or continuous system, but we are free to measure them directly. If our goal is to characterize $\chi$ itself, we want to work in a diagonal basis, 
\beq
\chi = \sum_{k=1}^K \nu_k |\lambda_k \rangle \langle \lambda_k |  \qquad \chi_N= \chi^{\otimes N}\ ,
\eeq
where $\lambda_k$ are the eigenvalues of some observable $\O$ that commutes with $\rho$, $[\O, \rho] =0$. Given our freedom to measure $\chi$ in any basis, we can measure $\O$ for each $\chi_i$ and calculate the probability that we measure a given $\lambda_k$ a fraction $n_k/N \to \mu(\lambda)$ of the time. Because we are working in a diagonal basis, this just follows Sanov's theorem with $\nu_k \to \nu(\lambda)$,
\beq\label{eq:diagsanov}
P(\{n_k \} ) = \exp\left(- N \sum_{\lambda \in \Lambda}^K \mu(\lambda) \log \frac{\mu(\lambda)}{\nu(\lambda)} \right) \ ,
\eeq
where $\Lambda = \{\lambda_k\}$ is the set of $\lambda$ values. This is not a quantum mechanical expression. Instead, we have exploited the classical nature of measurement in a single basis to use the classical theorem. Because we have assumed this is the diagonal basis, this is directly calculating the quantum relative entropy between $\chi$ and the empirical density matrix $\sigma  = \sum_k \mu_k |\lambda_k \rangle \langle \lambda_k|$ where $\mu_k = \mu(\lambda_k)$.

While we chose to diagonalize the density matrix $\chi$, the results didn't actually require it. Suppose that we are measuring $\O'$ without making any assumptions about the form for $\chi_N \equiv \chi^{\otimes N}$ in the $\O'$ basis. We want to know the probability that, after a sequence of measurements of the $\O'$ with eigenvalues $\lambda_k'$, the final state after measurement is a particular sentence
\beq
\Sigma(\{ \lambda_i' \}) = \sigma^{\otimes N} = \otimes_i (|\lambda'_{k_i,i} \rangle \langle \lambda'_{k_i,i} | ) \ .
\eeq
This is just the projection operator onto the state after the measurement. One can use this projection operator to define the \textit{quantum fidelity}, namely 
\beq\label{eq:fidelity}
F(\Sigma \| \chi_N) \equiv {\rm Tr} \sqrt{\Sigma} \chi_N \sqrt{\Sigma}  =\prod_{i=1}^N \chi(\lambda'_{k_i,i},\lambda'_{k_i,i}) \ .
\eeq
Note that we did not assume $\chi$ is diagonal, it is just that the repeated measuring in the $\O'$ basis picks out the diagonal elements. The quantum fidelity calculates the probability that we will be left with the density matrix $\Sigma$ after measurement given the initial density matrix $\chi_N$. If we performed a random walk in the $\O'$ basis, this probability is the same as calculating the probability of a specific random walk (i.e.~a unique sequence of steps). Instead, we want to eliminate the order of the individual measurement and calculate the probability of the histogram, $n_k/N \to \mu(\lambda')$. Including the combinatorial factor, $N!/(n_1! .. n_k!)$, the probability of the histogram is then
\beq
\log P(\{ n_k \}) = \log N! +\sum_{k=1}^K ( n_k \log \chi(\lambda'_k,\lambda'_k) - \log(n_k!) )  \ ,
\label{eq:QSanovIntermediate}
\eeq
where the sum over $k$ is the sum over the allowed eigenvalues, $\lambda'_k$, that leave the number of each bin, $n_k$, fixed. We notice this is actually related to the {\it measurement-induced relative entropy} between $\chi$ and an ``empirical" density matrix,
\beq
\log P(\{ n_k \}) = N D_{\lambda'}(\sigma \| \chi) \qquad \sigma = \sum_k \frac{n_k}{N} |\lambda'_k \rangle \langle \lambda'_k | \ ,
\label{eq:EmpiricalDensityMatrix}
\eeq
where $D_{\cal M}(\mu \| \chi)$ is the measurement-induced relative entropy in specific POVM basis ${\cal M}$~\cite{Berta:2017myy}.
This object appears in precisely the way we would expect from Sanov's theorem,
\beq
D_{\lambda'}(\sigma \| \chi) = D_{\rm KL}(\sigma_{\lambda'} \| \chi_{\lambda'}) \to P(\sigma) = e^{- N D_{\lambda'}(\sigma \| \chi)} \ ,
\eeq
where $D_{\rm KL}$ is the quantum relative entropy and $\sigma_{\lambda'},\chi_{\lambda'}$ are the density matrices after projective measurement in the $\lambda'$ basis, e.g. 
\beq
\chi_{\lambda'} = \sum_{k=1}^K |\lambda_k'\rangle\langle \lambda'_k | \langle \lambda'_k |\chi |\lambda_k'\rangle  \ .
\eeq
To some degree, this is just a quantum mechanical language for describing the classical relative entropy post measurement, as the density matrices become diagonal (classical) after decoherence. The purpose is that we will be able to generalize random walk or measurement of empirical distributions to any POVM description and the measurement-induced relative entropy will control the large deviation principle. 

The measurement-induced relative entropy appears precisely because it accounts for the projection of $\chi$ into the basis where the measurements occur. Quantum relative entropy of $\sigma$ and $\chi$, which appears in the quantum metrology definition of Sanov's theorem, does not depend on the basis in this way. As a result, the relative entropy will diverge when $\chi$ is a pure state (see Appendix~\ref{app:entropy} for further discussion).  This is indeed what happens in \cref{eq:diagsanov} when we are free to work in the basis that diagonalizes $\chi$. In the case quantum walk, the coupling of the walker to $\chi$ defines a preferred basis. The walk is only sensitive to specific matrix elements of $\chi$, that are coupled to the movement of our random walker. 

Having discussed the empirical density matrix for a measurements in a specific basis $\O'$, we now have the ingredients to reintroduce our walker. Rather than diagonalizing $\rho$, the probability of a given sequence of steps is determined by the Fidelity with respect the projector in the $s_i$ basis, 
\beq
\Sigma_s = \otimes_i (| s_i \rangle \langle s_i | ) \quad \to \quad \log P(\{s_i\}) =\log F(\Sigma_s \|\chi_N) \ .
\eeq
Similarly, the empirical distribution of steps is determined by measurement-induced relative entropy in the $s$ basis, 
\beq
 P(\sigma) = e^{- N D_{s}(\sigma_s \| \chi)} \qquad \sigma_s = \sum_{s\in {\cal S}} \frac{n_s}{N} |s \rangle \langle s | \ .
\eeq
It should be clear that we could generalize this statement to any quantum walk where the steps of the walk are i.i.d., in the sense that all steps are determined by a measurement of the same operator and identical density matrices $\chi$. Moreover, the generalization to a Markov chain where the steps and/or density matrices vary is still given in terms of $F(\Sigma \| \chi_N)$ where $\Sigma$ and $\chi_N$ are defined by the basis and density matrices that appear in each step of the walk.

The above construction assumes that we are measuring the individual steps of the walk in a fixed basis. Of course, we want to determine the probability of our walker ending at a particular value of $x$ after $N$ steps without making any measurements along the way. Now we see that this is achieved by minimizing the total measurement-induced relative entropy, just like the case of the classical random walk 
\bea
-\log P(x) =-\log \rho(x,x) &=& {\rm min}_{\rm s_i,x_0} ( -\log F(\Sigma_s \| \chi_N)- \log \rho_0(x_0,x_0) )\\
&=&  {\rm min}_{n_s,x_0}\left( N D_{s}(\sigma_s \| \chi)  - \log \rho(x_0,x_0)\right) \ ,
\eea
where the subscripts indicate the variables being minimized over (subject to the constraint from $x$).
This statement has a significant practical utility, given that both $\log \rho(x_0,x_0) \leq 0$ and $D_{s} \geq 0$.  For example, if we can calculate the WKB wavefunction to determine $\rho_0(x,x')$ but we are not able to determine the tail of the distribution of the walk itself, then we can still determine that 
\beq
P(x) \geq |\psi_{\rm WKB}(x)|^2 \ .
\eeq
We will see this will have useful applications in cosmology to determining lower bounds on the probability of large fluctuations.

Finally, notice that our discussion naturally generalizes to more complicated quantum walks. For example, even for identical $\hat \chi$ the walker can couple to different operators at each step. The walks we discuss only measure $\hat \chi_i$ in the position basis, $s_i$, but different steps could measure momenta, position, or any other hermitian operator. We can always describe a single walk in terms of the fidelity, \cref{eq:fidelity},  provided that we use the correct projection $\Sigma$.

\subsection{Classical Random Walks in a Potential}\label{sec:classical_stoch_with_pot}

The appearance of entropy in the description of i.i.d.~random walks suggests a deeper thermodynamic understanding of random walks more generally. In the rest of the section, we will review the idea of entropy production and rare events in simple non equilibrium settings, which is well established in the stochastic thermodynamics literature \cite{Seifert:2005rlb}. 

In the presence of external forces, it is less straightforward to describe the entropy, or free energy, of a single walker. The forces give rise to steps correlated in time, which makes the empirical distribution a less useful object. Working at large, $N \gg 1$, the most direct analogy with Sanov's theorem is to work the distribution defined by time averages of a single walker~\cite{donskervaradhan}. However, in the interest of clarity, we switch to describing the empirical distribution of large $M$ uncorrelated random walkers. Thus, in the remainder of this and the next section, we can think of the large deviation probability of releasing a swarm of $M$ walkers on a potential, and measuring their distributions. Moreover, the fact the $M$ walkers are uncorrelated means that we can interpret the results as a frequentist description of the probability distribution for a single walker.

Consider a Markovian process, the equilibrium state of which is $\pi(x)$. Under this dynamics, an initial distribution $P_0$ at time $0$ evolves to $P_f$ at time $\tau$. Focusing on a single particle that makes the journey from $P_0 \to P_f$, the entropy production associated with its path is the logarithm of the ratio of forward and reverse path probabilities (see \cref{app:Crooks}),
\begin{equation}
    s_{\rm tot}[x(t)] \equiv
    \log \left( \frac{P[x(t)]}{P[\tilde{x}(t)]} \right) \ . \label{eq:TotalEntropyPerPath}
\end{equation}
Here, $x(t)$ is the forward path that starts at $P_0$ and ends on $P_f$, and $\tilde{x}(t) = x(\tau-t)$ is the time reversed path that traverses backward.\footnote{\label{ft:Involution} When $x, p$ both are used, one should appropriately define time reversal as $x(t), p(t) \to x(\tau-t), -p(\tau-t)$, where $\tau$ is the ending time.} When detailed balance holds, this object reduces to (cf.\ \cref{eq:LogOfPathRatio})
\begin{equation}
    \log \left( \frac{P[x(t)]}{P[\tilde{x}(t)]} \right)
        = \log \frac{P_0(x_0)}{\pi(x_0)} - \log \frac{P_f(x_f)}{\pi(x_f)} \ ,
\end{equation}
where the distribution goes from $P_0 \to P_f$, both out of equilibrium.
Averaging over all paths, we obtain
\begin{align}
    D_{\rm KL} \left( P[x(t)] \| P[\tilde{x}(t)] \right)
        &= \int \mathcal{D} x P[x(t)] \log \frac{P[x(t)]}{P[\tilde{x}(t)]} \nonumber \\[0.5em]
        &= D_{\rm KL} ( P_0 \| \pi ) - D_{\rm KL} ( P_f \| \pi ) \equiv S_{\rm e} \ , \label{eq:ExcessEntropy}
\end{align}
where $S_{\rm e}$ stands for \textit{excess entropy}. Notably, it only depends on the initial and final snapshots $P_0$ and $P_f$, and the equilibrium distribution $\pi$ \cite{Vaikuntanathan2009}. Writing $\pi(x) = Z^{-1} \exp(-\beta V(x))$, we see that each KL term can be expressed in terms of free energies,
\begin{equation}
    D_{\rm KL} ( P \| \pi ) = - S[P] + \beta \langle U(x) \rangle_P + \log Z \equiv \beta({\cal F}[P] - {\cal F}[\pi]) \ . \label{eq:FreeEnergyExcess}
\end{equation}
Therefore, excess entropy is proportional to the drop in nonequilibrium free energy, $S_{\rm e} = \beta({\cal F}[P_0] - {\cal F}[P_f])$. 
Next, the probability that $M$ particles distributed as $P_f$ randomly fluctuate back to the initial state $P_0$ in time $\tau$ is

\begin{equation}
    \frac{\text{Prob}(P_f \to P_0)}{\text{Prob}(P_0 \to P_f)}
        = e^{-M S_{\rm e}} \ . \label{eq:ReversalProb}
\end{equation}
That is, the $P_f \to P_0$ transformation is exponentially unlikely compared to $P_0 \to P_f$, which is the Second Law of Thermodynamics. \cref{eq:ReversalProb} holds for any fixed $\tau > 0$ as $M \to \infty$, with the $\tau$-dependence contained within $S_{\rm e}$.
The forward relaxation $P_0 \to P_f$ is the native dynamics' typical behavior, so its probability is $O(1)$ and the ratio is governed entirely by the numerator. The most probable reversal is not an exotic excursion but the relaxation run backward \cite{Premkumar23}, so numerator and denominator compare a single trajectory against its own time-reversal: everything the dynamics does in between is shared between the forward and backward versions and cancels, leaving a pure boundary quantity, which is why the ratio depends only on $P_0$ and $P_f$ and not on the route. That boundary quantity is the free-energy drop. Free energy measures how far a distribution sits above equilibrium, so per particle, the price of being caught in an out-of-equilibrium configuration is exactly its free-energy excess over equilibrium (cf.\ \cref{eq:FreeEnergyExcess}). The difference in that excess between the two ends is what the reverse trip must fund. Since the $M$ particles are independent, reversing the whole swarm requires each to make its improbable return at once; the independent suppressions multiply and their exponents add, giving
$M$ copies of the per-particle cost.
This is manifest if we set $P_f = \pi$ in \cref{eq:ReversalProb} and use \cref{eq:ExcessEntropy}, which recovers Sanov's theorem:
\begin{equation}
    \text{Prob}(\pi \to P_0)
        \propto \exp(-M D_{\rm KL} \left( P_0 \| \pi \right)) \ .
\end{equation}
%

\begin{figure}[t]
    \centering
    \begin{subfigure}[b]{0.47\textwidth}
        \centering
        \begin{tikzpicture}[
            site/.style={circle, draw, thick, minimum size=1cm, fill=gray!10},
            fwd/.style={-{Latex[length=2.2mm]}, thick},
            bwd/.style={-{Latex[length=2mm]}, thin, gray}
        ]
            \node[site] (a) at (90:2.2)   {$a$};
            \node[site] (b) at (210:2.2)  {$b$};
            \node[site] (c) at (330:2.2)  {$c$};

            \draw[fwd] (a) to[bend left=15] node[pos=0.55, below right, black] {$p_{ab}$} (b);
            \draw[fwd] (b) to[bend left=15] node[midway, below, black] {$p_{bc}$} (c);
            \draw[fwd] (c) to[bend left=15] node[pos=0.6, left, black] {$p_{ca}$} (a);

            \draw[bwd] (b) to[bend left=15] node[midway, above left, gray] {$p_{ba}$} (a);
            \draw[bwd] (c) to[bend left=15] node[midway, below, gray] {$p_{cb}$} (b);
            \draw[bwd] (a) to[bend left=15] node[midway, above right, gray] {$p_{ac}$} (c);
        \end{tikzpicture}
        \label{fig:NESS_triangle_rates}
    \end{subfigure}
    \hfill
    \begin{subfigure}[b]{0.47\textwidth}
        \centering
        \begin{tikzpicture}[
            site/.style={circle, draw, thick, minimum size=1cm, fill=gray!10},
            current/.style={-{Latex[length=3mm]}, ultra thick, blue!70!black}
        ]
            \node[site] (a) at (90:2.2)   {$a$};
            \node[site] (b) at (210:2.2)  {$b$};
            \node[site] (c) at (330:2.2)  {$c$};

            \draw[current] (a) -- (b);
            \draw[current] (b) -- (c);
            \draw[current] (c) -- (a);

            \node[align=center, below=0.15cm] at (0,-1.2) {$J \neq 0$};
        \end{tikzpicture}
    \end{subfigure}
    \caption{A three-site lattice sustaining a nonequilibrium steady state (NESS). (Left) Individual transition rates to/from each site. Each bond has a dominant (thick) and a reverse (thin, gray) rate, with $\pi_a p_{ab} \neq \pi_b p_{ba}$ and cyclically for $(b,c)$ and $(c,a)$. (Right) The resulting net circulation. Since $\pi_a,\pi_b,\pi_c$ are constant in time, this current is stationary: mass leaving $a$ is exactly replenished via $c\to a$, with no accumulation anywhere. This persistent circulation, despite a perfectly stationary density, is the source of housekeeping entropy production.}
    \label{fig:NESS_triangle}
\end{figure}
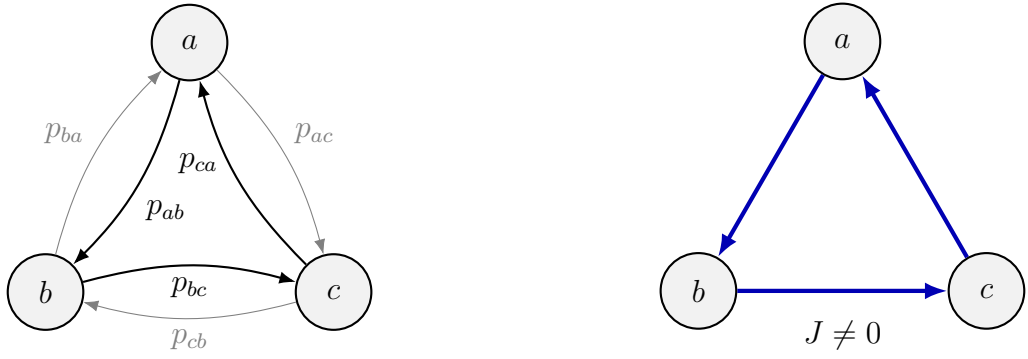

The arguments above can be extended to the case where detailed balance no longer holds \cite{Oono_Paniconi,Hatano2001}. That is, the probability mass moving back and forth between states do not balance each other out pairwise, even when the overall density is stationary. For example, consider a lattice with just three sites $a, b,$ and $c$, arranged as vertices of a triangle (see \cref{fig:NESS_triangle}). The probability mass that transfers from $a \to b$ in a given time step could be replenished by mass moving from $c \to a$ and so on. Thus there is a net circulation of probability mass in the triangle, even though snapshots of the probability distribution do not change from one instant to the next. This is called a Non-Equilibrium Steady State (NESS). We will denote it by $\pi_{\rm ss}$.

The probability current $J$ that sustains the NESS also picks out a preferred direction for the trajectories, even when the overall density remains steady. A path $x(t)$ that evolves along this direction is much more probable than its reversed version $\tilde{x}(t)$. This additional asymmetry between $P[x(t)]$ and $P[\tilde{x}(t)]$ is captured by the \textit{housekeeping entropy}
(see \cref{app:HousekeepigEntropy}),
\begin{equation}\label{eq:fluc_thm_NESS}
    \log \frac{P[x(t)]}{P[\tilde{x}(t)]} = \log \frac{P_0(x_0)}{\pi_{\rm ss}(x_0)} - \log \frac{P_f(x_f)}{\pi_{\rm ss}(x_f)} + s_{\rm hk}[x(t)] \ .
\end{equation}
Housekeeping entropy aggregates the departure from detailed balance at each point traversed by the particle (cf.\ \cref{eq:HousekeepingDefn}). Therefore, it is a path-dependent quantity. If we average over all paths, we obtain the \textit{total entropy} (cf.\ \cref{eq:ExcessEntropy})
\begin{equation}
    S_{\rm tot} \equiv D_{\rm KL} \left( P[x(t)] \| P[\tilde{x}(t)] \right) = S_{\rm e} + S_{\rm hk} \ , \label{eq:TotalEntropy}
\end{equation}
where $S_{\rm hk} \equiv \int \mathcal{D} x P[x(t)] s_{\rm hk}[x(t)]$, and $\pi$ is replaced with $\pi_{\rm ss}$ in the expression for $S_{\rm e}$. Total entropy measures the degree to which the system's dynamics is time irreversible. If $P_f = \pi_{\rm ss} = P_0$ entropy production is solely due to housekeeping. Then, the probability that the current runs in reverse under random fluctuations while the density stays at $\pi_{\rm ss}$ is proportional to $\exp(-M S_{\rm hk})$. Excess entropy is blind to the current reversal, as it only depends on the initial and final densities. When $P_f \neq \pi_{\rm ss} \neq P_0$, the overall probability of reversal in a system with a NESS is
\begin{equation}
    \text{Prob}[P_f \to P_0, J \to -J] \propto e^{-M S_{\rm tot}} . 
\end{equation}

As a simple example, consider a Fokker-Planck equation which becomes stationary ($\partial_t P = 0$), but has non-zero currents $(J \neq 0)$. For example, take a particle driven by a constant force $F$ and undergoing diffusion on a ring of circumference $L$. The position of the particle is governed by the Langevin equation $\dot{x} = \mu F + \sqrt{2g}\xi$. The associated Fokker-Planck equation is given by 
\begin{equation}\label{eq:classical_ring_FP}
    \frac{\partial P}{\partial t} = - \mu F \frac{\partial P}{\partial x} + g \frac{\partial^2 P}{\partial x^2} \ ,
\end{equation}
where $\mu$ is the mobility. Then, the stationary state is given by the uniform distribution $P(x) = 1/L$. However, even in this state we still have a constant current in the direction of the force,
\begin{equation}
    \partial_t P = - \partial_x J_x , \qquad J_x = \mu F P - g \frac{\partial P}{\partial x} = \frac{\mu F}{L} \ .
\end{equation}
The trajectories $x(t)$ overwhelmingly evolve in the direction of $J_x$, and housekeeping entropy captures the time irreversibility that arises from this preference. Then, the rate of entropy production in a NESS is given by the formula \cite{Seifert:2012pkl}
\begin{equation}\label{eq:hk_entropy_classical_ring}
    \dot{S}_{\rm tot} = \dot{S}_{\rm hk} = \int dx\,  \frac{J_{x}^2(x)}{g P(x)} \ ,
\end{equation}
which, for our example of the particle driven in a ring is gives $\dot{S}_{\rm tot} = \mu^2 F^2/g$. 

We will analyze \cref{eq:CosmoToyEOM} along the same lines in \cref{subsec:dSent}, but we can understand it qualitatively with the intuition developed above. Once again,
\begin{equation}
    \dot{x} = \frac{p}{m} + \sqrt{g} \xi \ , \qquad  \dot{p} = -\gamma p - V'(x) \ .
    \label{eq:CosmoToyEOM2}
\end{equation}
Notice that the damping term $\gamma p$ and the noise term $\xi$ are not co-located. This implies that time reversed version of certain trajectories are disallowed. To see why, consider a forward trajectory $(x(t), p(t))$ in phase space that evolves over the time interval $t \in [0,\tau]$. The time reversed trajectory $(\tilde{x}(t), \tilde{p}(t)) \equiv (x(\tau-t), -p(\tau-t))$ obey the equations (cf.\ \cref{ft:Involution})
%
\begin{equation}
    \dot{\tilde{x}} = \frac{\tilde{p}}{m} + \sqrt{g} \xi \ , \qquad  \dot{\tilde{p}} = \gamma \tilde{p} - V'(\tilde{x}) \ .
\end{equation}
The equations of motion of $p$ and $\tilde{p}$ are different; $\tilde{p}$ accelerates against friction! Therefore, all trajectories with $p(t) \neq 0$ at any $t$ have a time reversed version that is kinematically impossible, so $P[\tilde{x}, \tilde{p}] = 0$ even when $P[x, p] \neq 0$. Then, the KL between these probabilities diverge, making $S_{\rm tot} = \infty$. Furthermore, it can be shown that this infinity resides in the housekeeping entropy, since $S_{\rm hk}$ is the term that picks up time irreversibility at the level of trajectories (see \cref{subsec:dSent}).

It is worth noting that, if we add a noise term to the $p$ equation
\begin{equation}
    \dot{p} = - \gamma p - V'(x) + \sqrt{2 g_p} \xi_1 \ , \qquad \dot{\tilde{p}} = \gamma \tilde{p} - V'(\tilde{x}) + \sqrt{2 g_p} \xi_1 \ .
\end{equation}
Then, the anti-damped trajectory can be attributed to the noise, so the time reversed trajectory is assigned a probability $\propto \exp[- \frac{1}{4 g_p^2} \int d t (\dot{\tilde{p}} - \gamma \tilde{p} + V'(\tilde{x}))^2]$. Then, the KL between forward and reverse path measures is well-behaved, and $S_{\rm tot}$ becomes finite. The noise on $x$ in \cref{eq:CosmoToyEOM2} cannot regulate $S_{\rm tot}$ this way because it must account for the reverse fluctuations of $x$ itself. Once any $x$ path is observed the $p$ path is determined exactly by integrating $\dot{p} = -\gamma p - V'(x)$, but the corresponding $\tilde{p}$ is unphysical in the absence of $\xi_1$.

\subsection{Quantum Walks with the Potential}\label{sec:quantum_stoch_with_potential}

We would like to explain the quantum analogue of the ideas introduced in \cref{sec:classical_stoch_with_pot}. We will again use the idea of large $M$ walkers to understand the large deviation probability of various observables. Since the walkers evolve independently, this can be understood as giving a frequentist definition of the probability after a given set of quantum measurements. It will also be important to note that in quantum systems, the Fokker-Planck equation is satisfied by the Wigner function, which is not a probability distribution. This will mean that we cannot think of $-\int W(x_c,p_q)\log W(x_c,p_q)$ as the entropy of a distribution. Rather, only after marginalizing over one variable do we recover the probability interpretation. This will have an important effect when we discuss the housekeeping entropy in the quantum theory, which classically required an analysis of the path level entropy of the full phase space. 

We first consider the case of the quantum thermal system, where we take $M$ quantum particles in the thermal density matrix 
\begin{equation}
    \rho = \frac{\exp(-\beta \hat{H})}{\Tr(\exp(-\beta \hat{H}))} \qquad \chi_N = \rho^{\otimes M} \ .
\end{equation}
Now, we want to measure in a given basis, say $x$. Define $\sigma(x)$ as the ``empirical" density matrix, defined by the distribution of measurements in the $x$-basis (cf.\ \eqref{eq:EmpiricalDensityMatrix}) 
\beq
\sigma = \int dx Q(x) |x \rangle  \langle x | \ ,
\eeq
where we wrote a continuous distribution $n_x/M \to Q(x)$ as we did in the classical theory. We want to find the probability of observing $Q(x)$ given the true equilibrium density matrix $\rho$. Using the quantum Sanov's theorem, we get that the probability to find $\sigma_x$ is determined by 
\begin{equation}
  M D_{x}  = M\langle \log \sigma\rangle_\sigma -M\beta\langle H\rangle_\sigma = M \beta {\cal F}(Q) \ .
\end{equation}
which is the free energy associated with the distribution $Q(x)$, the diagonal elements in the $\sigma$ distribution, and $\langle .. \rangle_\sigma$ means the average calculated using the density matrix $\sigma$. Thus, in the case of $M$ quantum particles in equilibrium, the measurement-induced relative entropy of the distribution determines the cost of atypical fluctuations: 
\begin{equation}\label{eq:eq_free_en}
    P(\sigma)_{\rm QM} = \exp(-M D_{x}(\sigma \|\rho)) \ .
\end{equation}
This result is for the static case, where we are essentially picking out particles from an i.i.d~distribution. Now, we want to extend to the dynamical case, where the density matrix evolves according to some Lindbladian. We again start with the entropy production formula of a path as
\begin{equation}
    s_{\rm tot} = \log \left ( \frac{P[x(t)]}{P[\tilde{x}(t)]} \right) \ .
\end{equation}
For a quantum theory, a path is not a well defined quantity. One protocol to follow is to measure a path transition by looking at the jump statistics of the Lindblad operators \cite{Horowitz:2013mfk, Manzano:2015rnk}. As a simple example, if we flip coins for each step for our quantum walker, rather than measuring the position of the walker, we just look at the values of the flipped coin measurements. This definition does not affect the evolution of the walker because we end up tracing over the coins anyway. This, combined with making two measurements of the walker at the initial and final times, lets us define a forward and a reverse trajectory. Then, for the detailed balance case, we recover
\begin{equation}
    s_{\rm tot} = \log \frac{P_0(x_0)}{\pi(x_0)} - \log \frac{P_f(x_f)}{\pi(x_f)} \ . 
\end{equation}
Using this protocol, we see that similar to the classical case, assuming that the optimal path for the $P_f \to P_0$ is given by the time reverse of $P_0 \to P_f$, we recover that 
\begin{equation}
    \frac{\text{Prob}(P_f \to P_0)}{\text{Prob}(P_0 \to P_f)}
        = e^{-M S_{\rm e}} \ . 
\end{equation}
which is just the quantum generalization of \cref{eq:ReversalProb}. Similarly, one can find the housekeeping entropy using a similar analysis \cite{Horowitz:2013mfk, Manzano:2015rnk}.

However, this is an indirect measurement of the quantum path, which is not a priori always possible to setup. For instance, in cosmology, when a light field undergoes stochastic inflation, the observer does not have the statistics of the Lindblad operators available to measure. Thus we want to extend our discussion to the case where we only have the random walker available to measure. Since measuring paths in quantum mechanics leads to wave function collapse along the observable, it will be useful to build intuition for the quantum theory with a simple example. 

Analogous to the classical case we consider in \cref{eq:classical_ring_FP}, we can construct a model for a quantum walker where the walker lives on a ring, and is pushed by a constant force $F$, to give the Langevin equation 
\begin{equation}
    \dot{x}_c = \mu F + \xi \ .
\end{equation}
Such a walker can be constructed explicitly via 
\begin{equation}
    \rho(x,x') = \bigg\langle x \bigg| {\rm Tr}_i \exp \left(i \sum_i \hat p (\hat{s}_i + \mu F) \right) \rho_0 \otimes \chi^{\otimes N}  \exp \left(-i \sum_i \hat p (\hat{s}_i + \mu F) \right) \bigg| x' \bigg\rangle \ ,
\end{equation}
where at each time, we introduce noise via $s_i$ and a constant push in the $+x$ direction via $\exp(i \hat{p} \mu F)$. The evolution of this density matrix is then given by 
\begin{equation}
    \frac{\partial \rho}{\partial t} = - \mu F \left(\frac{\partial}{\partial x} + \frac{\partial}{\partial x'} \right)\rho  + \frac{g}{2} \left( \frac{\partial}{\partial x}  + \frac{\partial}{\partial x'}\right)^2 \rho  \ .
\end{equation}
Now, on measuring the $\hat{x}$ direction of such a walker, we will reduce to the classical particle on a ring
\begin{equation}
    \frac{\partial P(x,t)}{\partial t} = - \mu F\frac{\partial P(x,t)}{\partial x} + \frac{g}{2}\frac{\partial^2 P(x,t)}{\partial x^2} \ ,
\end{equation}
and we would see constant housekeeping entropy production as given in \cref{eq:hk_entropy_classical_ring}. But of course, since we are completely agnostic to any open system dynamics in the momentum $p$ direction, if we were to measure the $p$ distribution, we would see that 
\begin{equation}
    P(p,t)  = P(p, t_0)
\end{equation}
so that the distribution is constant in time, and there is no entropy production. This illustrates how when we project onto a certain measurement basis in quantum mechanics, we can hide the entropy production due to the NESS. 

Thus, while $\rho(x,x',t)$ follows a Markovian Lindblad equation, we want to more accurately describe the entropy production related to the variable we want to measure. This is similar in spirit to the quantum Sanov's theorem, where we use the measurement-induced relative entropy to ultimately project onto the basis we are measuring in. Alternatively, as mentioned in the beginning of the section, the phase space distribution is given by the Wigner function, which is a quasi-probability distribution, and can take negative values. To get well defined entropy, we have to marginalize the Wigner function and get a probability distribution of the variable we are measuring. Thus, given the variable $a = f(x,p)$, whose basis we are measuring in, formally we define the entropy production as 
\begin{equation}
    s_{\rm tot}[a(t)] = \log \left(\frac{P[a(t)]}{P[\tilde{a}(t)]} \right) \ ,
\end{equation}
where $P[a(t)] = \bra{a}\hat{\rho} (t)\ket{a}$ is the path level probability distribution for the variable $a = f(x,p)$ which one gets from the density matrix. The important caveat to this equation, however, is that the evolution of $P[a(t)]$ is no longer required to be Markovian, since we are integrating out the conjugate variable. 

Sometimes, as in the example above of the particle diffusing on the ring, when we marginalize the distribution and find the evolution just for the distribution of $a$, $P(a,t)$, we will get back a Markovian Fokker-Planck equation, in which case we can use our classical results from the previous subsection to analyze entropy production. One can also use approximate equations, as in overdamped systems, to systematically eliminate a fast variable and get back a Markovian evolution for the distribution of the slow variable. In both these cases, one can analyze the housekeeping entropy of this effectively classical one-dimensional Fokker-Planck equation, and recover the same fluctuation theorems \cref{eq:fluc_thm_NESS}. However, due to non-Markovian effects, in general one has to explicitly analyze the resulting evolution equations to extract the total entropy. 

The central conclusion from this section is that we can apply the same thermodynamic understanding of quantum walks as we did for classical walks. Quantum mechanics does not fundamentally change the nature of the evolution itself. However, when applied at the level of the Wigner function, the observability of the housekeeping entropy is not a given. The measurements of $x_c$ and $p_q$ do not commute, meaning that the evolution of the system needs to account for the measurement process. For simplicity, we have focused only on measurements in a single fixed basis, but we will investigate these more complex measurements in future work.

\section{Cosmological Applications}\label{sec:cosmology}

Cosmology is a unique setting where both classical statistical physics and quantum mechanics play essential roles. Structure in the Universe is widely believed to have originated from quantum vacuum fluctuations~\cite{Green:2020whw}. At the same time, the long wavelength universe behaves mostly like a classical stochastic system, where the quantum fluctuations are the source of noise. However, rare fluctuations in this context can literally reshape the entire universe, from generating primordial black holes~\cite{Carr:2025kdk} to driving eternal inflation~\cite{Linde:1986fd,Goncharov:1987ir}. Prior work has shown why these calculations are difficult to control. Our improved understanding of quantum tails now enables us to give a more precise articulation of situation. We will derive the relationship between stochastic inflation and the WKB approximation in Section~\ref{subsec:stochastic} and then apply it to the DBI action in Section~\ref{subsec:dbi}.

Naturally, a deeper understanding of the role of entropy in quantum stochastic processes provides a stepping stone to further connecting stochastic inflation~\cite{Arkani-Hamed:2007ryv} and the Gibbons-Hawking entropy~\cite{Gibbons:1977mu}. In Section~\ref{subsec:dSent}, we will show that the quantum generalization of stochastic inflation does not reach detailed balance and leads to violations of the KMS condition.

\subsection{Stochastic Inflation versus the Wavefunction}\label{subsec:stochastic}

Calculating rare fluctuations in cosmology presents an apparent contradiction between two arguments, one favoring WKB and the other favoring stochastic inflation. Although we can anticipate that the relative importance will follow a pattern similar to Section~\ref{sec:qw}, understanding how these arguments fail will be important as well.

\begin{figure}[!ht]
    \centering
    \includegraphics[width=0.85\textwidth]{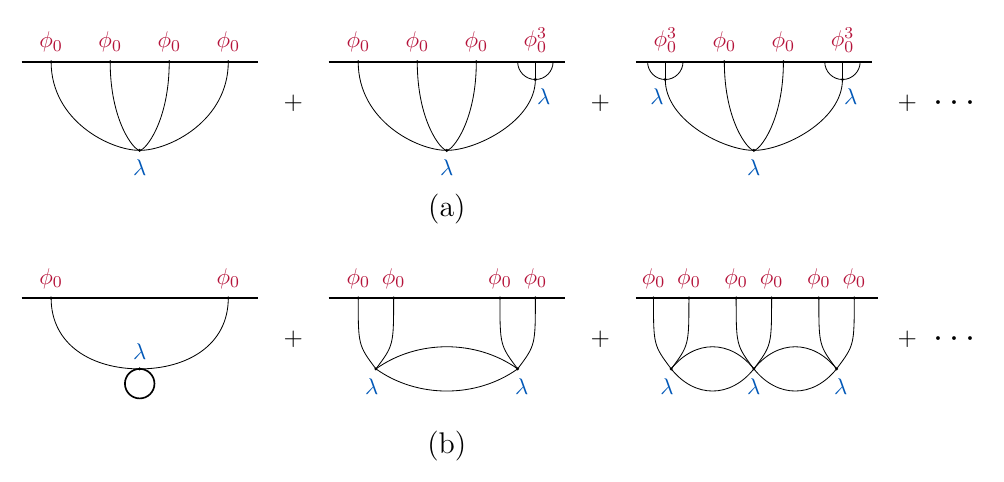}
    \caption{Contributions to the wavefunction from (a) tree level and (b) loop-level Witten diagrams. The tree level diagrams dominated when $\phi_0^2\lambda \gg 1$ as the loop diagrams are suppressed by additional powers of $\lambda$ for the same order in $\phi_0$. Figure adapted from~\cite{Celoria:2021vjw}.
    }
    \label{fig:wavefunction}
\end{figure}

Quantum fluctuations in cosmology are often described in terms of a wavefunction. In quantum mechanics, we are used to the idea that the tail can be calculated using the WKB approximation. In~\cite{Celoria:2021vjw}, it was argued that rare fluctuations in cosmology are determined by the semi-classical saddle which determines the ground state wavefunction. In the context of QFT or perturbative quantum gravity, the argument is based on power counting loop diagrams, and is illustrated in Figure~\ref{fig:wavefunction}. This scaling behavior of diagrams is known to explain why semi-classics control scattering amplitudes at large multiplicity~\cite{Voloshin:1992mz}.

To understand the idea, we remember that perturbative evaluation of the wavefunction of the Universe for $\phi$ can be organized in terms of Witten diagrams. However, rather than using it to isolate some correlators, we want to calculate the leading contribution to $\Psi[\phi_0]$ when $\phi_0 \gg H$. The idea is that we can use scaling to isolate a subset of the Witten diagrams that are responsible and determine their behavior non-perturbatively. The key observation is that only bulk-to-boundary propagators directly depend on the value of $\phi_0$. As a result, we would expect the powers of $\phi_0$ should be fixed only in terms of the number of bulk to boundary legs.

Suppose, for example, we have an interaction $V(\phi) =\frac{1}{4!} \lambda \phi^4$. For every interaction vertex that only involves bulk-to-boundary propagators, we will find a contribution to the wavefunction proportional to $\lambda \phi_0^4$. For tree-level diagrams, each new vertex must involve one bulk-to-bulk correlator, which removes two factors of $\phi_0$, one for each end. As a result, for tree-level, the scaling behavior of the wavefunction when $\phi_0^2 \lambda \gg 1$ is given by
\beq
\log \Psi_{\rm tree}[\phi_0] \supset H^{-2} \phi_0^2 \sum_n c_n (\lambda H^{-2} \phi_0^2)^n \ .
\eeq
 Now we introduce loops, such as those in the lower panel of Figure~\ref{fig:wavefunction}. Since the loops always involved bulk-to-bulk propagators, we only find additional powers of $\lambda$ with no additional factors of $\phi_0$. As a result, the wavefunction when $\phi_0^2 \lambda \gg 1$ is dominated by the semi-classical action (the sum of the tree diagrams)
 
 However, a critical assumption in this argument is that the loop corrections are suppressed by $\lambda$ (potentially after renormalization). When this assumption is correct, it is indeed true that the diagrams in (a) dominate the wavefunction and correspond to a semi-classical solution. This is the quantum field theory analogue of the WKB solution in \cref{eq:WKB}. As a result, the density matrix is a pure state, like \cref{eq:WKB_rho}, $\rho =|\Psi \rangle \langle \Psi |$ so that 
\beq
\dot \rho = -i [ H, \rho] \ .
\eeq
The analytic solution for WKB is then accurate to all-orders in $\lambda \phi_0^2$ at leading order in $\lambda$.

The problem with this argument is that the loop correction are often infinite. Of course, we can regulate the divergences so that these results are finite but may include large logs that need to be resummed by RG. This is precisely what happens for light fields in de Sitter~\cite{Burgess:2014eoa,Burgess:2015ajz,LopezNacir:2016gzi,Gorbenko:2019rza,Baumgart:2019clc,Mirbabayi:2019qtx,Cohen:2020php,Mirbabayi:2020vyt,Baumgart:2020oby,Palma:2023uwo,Beneke:2023wmt,Palma:2025oux,Beneke:2026rtf,Beneke:2026ksj} and therefore we cannot use the scaling with $\lambda$ alone to determine the leading behavior. Instead, we also have a subset of diagrams that scale as $\lambda \log k/(a(t)H)$, where $a(t) = e^{Ht}$ is the scale factor and $k\ll aH$ is a super-horizon fourier mode. Without understanding how to resum these diagrams as $-\log k/aH \to \infty$, we cannot conclude that the loops are sub-dominant. 

Using the methods of exact RG, one finds that the full problem is identical to the walk we described in Section~\ref{sec:noise_in_x}. Integrating out a shell of momentum creates a mixed state for a local measurement of $\phi$~\cite{Li:2025azq,Green:2025hmo}, giving rise to an open quantum system. The evolution is most easily described in terms of the Wigner distribution, $W(\bar \phi, q)$, which to leading order is
\beq
\frac{d}{d t} W[\bar{\phi}, q]=\frac{H^3}{8 \pi^2} \frac{\partial^2 W}{\partial \bar{\phi}^2}+3 H \frac{\partial}{\partial q}(q W)-q \frac{\partial}{\partial \bar{\phi}} W+\frac{\partial}{\partial q} V^{\prime}(\bar{\phi}) W +\ldots 
\eeq
The terms with $\ldots$ involve higher powers of the fields and/or derivatives. Ignoring these higher order terms and integrating out $q$ gives rise to stochastic inflation for the classical probability distribution, $P(\bar \phi)$.

If we focus specifically on the case of $V(\phi) = \frac{1}{4}\lambda \phi^4$, then one can check explicitly that the Stochastic formalism gives the dominant contribution to $\rho$ when $\lambda \phi_0^2 < 1$. Solving the Fokker-Planck equation gives the familiar equilibrium probability distribution
\beq
P(\bar \phi) = c\exp\left(- \frac{3V(\bar \phi)}{8 \pi H^4}\right) \ .
\eeq
This is precisely the same result as we get for a quantum walk where classical noise dominates, \cref{eq:thermal_eq}.

So far, there is no contradiction because the regimes of validity between the dominant contributions do not overlap. However, the stochastic description allows us to calculate corrections as an expansion in $\lambda\phi_0^2/H^2$ and it is not true that they are always sub-dominant to WKB at large $\phi_0$. Given that probable values of $\phi_0$ are bounded by $|\phi_0|\lesssim\lambda^{-1/4}$, this appears to be an expansion in $\sqrt{\lambda}$ for many statistical averages. These corrections have a natural interpretation in terms of the Kramers-Moyal expansion of a general Markovian random walk~\cite{Cohen:2021fzf}
\bea
\frac{\partial}{\partial t} P(\phi, t)&=&\int \mathrm{d} \Delta \phi[P(\phi-\Delta \phi, t) \widetilde{W}(\Delta \phi, \phi-\Delta \phi)-P(\phi, t) \widetilde{W}(\Delta \phi, \phi)] \\
&\to&
\sum_{n=1}^{\infty} \frac{1}{n!} \frac{\partial^n}{\partial \phi^n} \Omega_n(\phi) P(\phi, t) \ ,
\eea
where the higher derivatives encode the non-Gaussian moments of the transition amplitude, 
\beq
\Omega_n(\phi) \equiv \int \mathrm{d} \Delta \phi(-\Delta \phi)^n W(\phi+\Delta \phi \mid \Delta \phi) \ .
\eeq
For example, in $\lambda \phi^4$, it was found in~\cite{Cohen:2021fzf} that $\Omega_3 \propto \lambda \phi$, corresponding to the skewness of the steps of a random walk.

Naively, the breakdown of the Kramers-Moyal expansion when $\lambda \phi^2_0\gg 1$ would seem to be consistent with the dominance of the WKB solution. While it is true that the Kramers-Moyal expansion fails in the same regime that quantum effects become important~\cite{Cohen:2021jbo,Cohen:2022clv}, it does not follow that the quantum effects dominate in this regime. Instead, both arise from the fact that the tail of the distribution is sensitive to the non-perturbative microphysics of the theory and therefore cannot be reliably calculated within perturbation theory~\cite{Cohen:2021jbo,Cohen:2022clv}. WKB offers a non-perturbative description of the vacuum state, but neglects the non-perturbative behavior of the noise generated by the short distance fluctuations.

In short, we cannot calculate the transition amplitude $W(\phi+\Delta \phi|\Delta \phi)$ within effective theory when the size of the jumps, $\Delta \phi$, become large. Only the full non-perturbative definition of the theory can provide the behavior in this regime. WKB effectively sets $\Delta \phi=0$ and determines only the quantum contribution to the wavefunction.

In order to understand the challenge in more detail, suppose we have a quantum walk given by the Hamiltonian
\beq
H=\frac{p^2}{2 m}+V(x)+p \sum_i s_i .
\eeq
where $s_i$ are additional quantum mechanical degrees of freedom with some density matrix $\chi_i$, that we will assume is the same for all $i$. The first two terms give our quantum evolution and the last term is our classical noise that could be drawn from any distribution.
Following~\cite{Green:2025hmo}, the evolution equation Wigner distribution,
\beq
W(\bar x,p) =  \int d\Delta x \rho_R(\bar x-\Delta x/2, \bar x +\Delta x/2)e^{i p \Delta x} \qquad \rho_R ={\rm Tr}_{s_i} \rho(x,x'; \{ s_i, s_i' \}) \ , 
\eeq
the evolution equation takes the form
\beq
\frac{\partial W}{\partial t}=-i\left(V\left(x+\frac{i}{2} \partial_p\right)-V\left(x-\frac{i}{2} \partial_p\right)\right) W-\frac{p}{M} \frac{\partial W}{\partial x} +\sum_{n\geq 2} \frac{\gamma_n}{n!} \frac{\partial^n W}{\partial x^n}.
\eeq
where the first two terms are the Hamiltonian evolution of the quantum system and $\gamma_n$ are the cumulants of the distribution $\chi(s_i,s_i)$ that represent the classical noise.

If we want to calculate $W(x,p)$ at large $x$, we need to understand how the infinite series in $\gamma_n$ compares to the Hamiltonian evolution, which requires a non-perturbative analysis of the microphysics. Fortunately, because the probability of rare fluctuations is determined by minimizing over both quantum and classical fluctuations, we can still determine a lower bound on the probability distribution from the WKB solution alone.

\subsection{Bound on Tails in DBI Inflation}\label{subsec:dbi}

Although the UV calculation of the classical noise remains beyond the scope of current techniques, the min-relative entropy provides a lower-bound on the tail of the distribution from the wavefunction alone. For scalar fields with canonical kinetic terms and a potential $V(\phi)$, the result can be determined using the WKB description in Section~\ref{sec:entropy}. We therefore turn our attention to theories with non-trivial interactions that have a known UV description. For this reason, DBI inflation is a natural model in which to investigate non-trivial tails.

The DBI action for a scalar field with a potential $V(\phi)$ is given by~\cite{Silverstein:2003hf,Baumann:2006cd}
\beq
S = \int d^4 x \sqrt{-g} \left(-\frac{1}{f(\phi)}\sqrt{1- f(\phi) \partial_\mu \phi \partial^\mu \phi} + \frac{1}{f(\phi)}- V(\phi) \right)  \ ,
\eeq
where $g$ is the metric and $f(\phi)$ is a generic function that arises from the warp factor in higher-dimension brane-inflation models. We recover the action of a canonical scalar field in the limit $f(\phi) \to 0$. 

In order to understand the tail of the distribution for a the measurement of $\phi(\x=0,t) \equiv \varphi(t)$ at a single-point $\x = 0$, we will trace over both UV modes, $k>aH$ and the field values at separated points $\phi(\x \neq 0 , t)$. This tracing out procedure was demonstrated in~\cite{Green:2025hmo} and gives rise to a $0+1$d a Markovian (Lindbladian) evolution equation for the density matrix $\rho(\varphi,\varphi')$,
\beq
\dot \rho = -i [H, \rho] + \sum_i \left(L_i \rho L_i^{\dagger}-\frac{1}{2}\left\{L_i^{\dagger} L_i, \rho\right\}\right) \ .
\eeq
The key observation here is that $H$ is the 0+1 dimensional Hamiltonian for the unitary evolution of $\varphi(t)$. Because the theory becomes ultra-local in the limit of large wavelength, this effective Hamiltonian arises from the UV Hamiltonian projected to $\x=0$.

Our goal here is not to determine $\rho(\varphi, \varphi')$ but to determine the lower bound on $P(\varphi) =\rho(\varphi,\varphi)$ in the regime $\lambda \varphi^2 \gg 1$ for $V(\phi) = \lambda \phi^4$. Since this is now an effectively quantum mechanics problem, we only need to determine the WKB wavefunction for 
\beq
S_{0+1d,\pm} = \int_{-\infty(1\pm i \epsilon)} dt \left(-\frac{1}{f(\varphi)}\sqrt{1- f(\varphi) \dot \varphi^2} +\frac{1}{f(\varphi)}- V(\varphi) \right) \ .
\eeq
This should hold in the limit where the classical noise, generated by the jump operators $L_i$ vanishes. This can be best understood in the Schwinger-Keldesh formalism as follows: the semi-classical density matrix is given by 
\beq
\rho(\varphi, \varphi') = \exp\left( i S_{0+1d,+}[\varphi] -i S_{0+1d,-}[\varphi'] + i S_{\rm noise}(\varphi,\varphi')  \right) \ .
\eeq
Although we have the exact form the the kinetic term, the calculation of the noise non-perturbatively is required to determine the tail. However, provided the noise vanishes along the path, $S_{\rm noise}= 0$ when $\varphi(t') = \varphi'(t')$ and $\dot \varphi(t') = \dot \varphi'(t')$ for $t' \in (-\infty(1\pm i \epsilon), t]$, then we can 
\beq
P(\varphi) \approx \exp\left( -2{\rm Re} S_{0+1d}[\varphi] \right) \ .
\eeq
The real action here is the same as the WKB calculation described in Section~\ref{sec:qw} but not with $\beta \to \infty$.

The easiest way to calculate the exponentially suppressed contribution at large $\varphi$ for the ground-state wavefunction is using the continuation to Euclidean signature:
\beq
S_{E} = \int dt_E \left(\frac{1}{f(\varphi)}\sqrt{1+ f(\varphi) \varphi'{}^2} - \frac{1}{f(\varphi)}+ V(\varphi) \right) \ ,
\eeq
where $\dot \varphi = \partial_{t_E} \varphi(t_E)$, much like the case of tunneling in DBI~\cite{Brown:2007zzh,Sarangi:2007jb}. The saddle point approximation will give a good approximation to the path integral when $S_{E} \gg 1$ (in units where $\hbar =1$). We will find the saddle point starting from the conjugate momentum and Hamiltonian, 
\bea
p&=& \frac{\varphi'}{\sqrt{1+ f(\varphi) \varphi'{}^2}} \to \varphi'{}^2 = \frac{p^2}{1-f(\varphi)}\ , \\
H_E = p \varphi' - {\cal L}_E 
&=& \frac{1}{f} \left(1- \frac{1}{\sqrt{1+f \varphi'{}^2}} \right)-V(\varphi) \ .
\eea
Since energy is conserved, $H_E = {\rm constant}$. Since we are in the ground state, we take $H_E = 0$ to find 
\bea
H_E= 0 \quad \to \quad \frac{1}{\sqrt{1+f \varphi'{}^2}} = 1- f(\varphi) V(\varphi) \\
f \varphi'{}^2 = \frac{fV(2 -fV)}{(1-f V)^2} \quad \to \quad \  p =  \sqrt{V(\varphi) (2-f(\varphi)V(\varphi))} \ .
\eea
We can now calculate the Euclidean action,
\beq
S_E =\int dt_E L = \int dt_E (p(\varphi) \phi' -H_E) = \int_{\varphi_0}^{\bar \varphi} d\varphi \sqrt{V(\varphi) (2-f(\varphi)V(\varphi))} \ ,
\eeq
where $\varphi_0$ is the minimum of the potential $V(\phi)$. The resulting wavefunction is given by
\beq
\Psi(\bar \varphi) \approx \exp(-\int_{\varphi_0}^{\bar \varphi} d\varphi \sqrt{V(\varphi) (2-f(\varphi)V(\varphi))}  \ .
\eeq
This behavior holds as long as $S_E \gg 1$, which fails as $f(\varphi) V(\varphi) \to 0$. 

When $f(\varphi)V(\varphi) >2$, the action is purely imaginary and therefore not well described by the saddle point approximation. This regime can be understood as parameters space where the Brown-Teitelboim~\cite{Brown:1988kg} mechanism for pair production becomes allowed by energy conservation. As we are working with a single worldline, we can see this more easily using a conventional massive particle action
\beq
S=\int d\tau\left(- m \sqrt{1-x'^2}+m -V(x)\right) \ .
\eeq
We see that $f(\phi) \to 1/m$ and $\dot x^2 = \dot \varphi^2/f$ relates these two descriptions. As a result, the Euclidean instanton no longer gives a reliable approximation when
\beq
f(\varphi)V(\varphi) > 2  \to V(x) > 2 m \ .
\eeq
This is consistent with the explanation that it is the onset of pair production that is responsible for the change in behavior of the Euclidean solutions.

\subsection{Stochastic Inflation and the de Sitter Entropy}
\label{subsec:dSent}

Some of the most pressing questions in both theoretical cosmology and quantum gravity~\cite{Flauger:2022hie} are the interpretations of the Gibbons-Hawking entropy~\cite{Gibbons:1977mu} and the Hartle-Hawking wavefunction of the Universe~\cite{Hartle:1983ai}. At the same time, understanding the natural of quantum fields in de Sitter has lead to the idea that the local physics is equivalent to a random walk~\cite{Maldacena:2024uhs}. This formalism, known as stochastic inflation~\cite{Vilenkin:1983xq,Starobinsky:1986fx,Starobinsky:1994bd} has since been derived directly from QFT~\cite{Burgess:2014eoa,Burgess:2015ajz,LopezNacir:2016gzi,Gorbenko:2019rza,Baumgart:2019clc,Mirbabayi:2019qtx,Cohen:2020php,Mirbabayi:2020vyt,Baumgart:2020oby,Palma:2023uwo,Beneke:2023wmt,Palma:2025oux,Beneke:2026rtf,Beneke:2026ksj} and has been extended to a quantum walk for the Wigner distribution function~\cite{Nambu:1991vs,Li:2025azq,Green:2025hmo,Calderon-Figueroa:2025dto,Cespedes:2026fdp,Li:2026lwl,Christie:2026dwx}. Naturally, one would hope to connect the quantum walk with the de Sitter entropy and the Hartle-Hawking wavefunction (or density matrix~\cite{Page:1986vw,Hawking:1986vj,Ivo:2024ill}).

For a field with no potential is dS, it has long been known that the random walk description generates an infinite amount of entropy. The random walk is unbounded, and the number of modes that are frozen diverges. This is consistent with quantum gravity in dS because the dS entropy also diverges as $\Mpl\to \infty$. The natural question is what happens at finite $\Mpl$. One might try to address this question using inflation~\cite{Arkani-Hamed:2007ryv} and it was argued that slow-roll eternal inflation prevents us from observing more modes than the de Sitter entropy. However, this conclusion depends on the calculation of the tail of the distribution and is generally not under calculable with current techniques~\cite{Cohen:2021jbo}.

It is also natural to suspect that quantum gravity prevents us from writing a scalar field theory without a potential. A confining potential causes the theory to reach equilibrium and therefore one might have assumed entropy production halts. However, as shown in~\cite{Nambu:1991vs,Li:2025azq,Green:2025hmo,Calderon-Figueroa:2025dto,Cespedes:2026fdp,Li:2026lwl,Christie:2026dwx}, at the level of the density matrix, stochastic inflation does not reach detailed balance and leaves open the possibility of infinite housekeeping entropy even when $V(\phi) \neq 0$.

As we see in Section \ref{sec:classical_stoch_with_pot},  a NESS has a constant entropy production even at equilibrium. In \cite{Green:2025hmo}, it was found that the phase space evolution for stochastic inflation is given via the Langevin equations
\begin{equation}
    \dot{x}_c = \frac{p_q}{m} + \sqrt{g}\xi \qquad \dot{p}_q = -V'(x_c) - \gamma p_q \ .
\end{equation}
As was already noted, such a distribution does not follow detailed balance at equilibrium, but rather represents a NESS. As we have discussed, even at equilibrium, a NESS constantly produces entropy, called the housekeeping entropy. In this section we want to explicitly calculate the housekeeping entropy production for the case of a harmonic force (a light mass) in stochastic inflation.

The Fokker-Planck equation for stochastic inflation is given by the evolution for the Wigner function
\begin{equation}
    \frac{dW}{dt} = \frac{g}{2} \frac{d^2 W}{dx_c^2} - \frac{p_q}{m} \frac{d W}{dx_c} + \frac{d}{dp_q}(\gamma p_q W) + i\left(V\left(x_c + \tfrac{i}{2} \partial_{p_q}\right)-V\left(x_c - \tfrac{i}{2} \partial_{p_q}\right)\right) W \ .
\end{equation}
with $g = H^3/4\pi^2, \gamma = H$. For the case of $V'(x) = kx$, we can expand out the potential to get
\begin{equation}
    \frac{dW}{dt} = \frac{g}{2} \frac{d^2 W}{dx_c^2} - \frac{p_q}{m} \frac{d W}{dx_c} + \frac{d}{dp_q}(\gamma p_q W + V'(x_c) W) \ .
\end{equation}
Since this is Gaussian, the Wigner function will be positive definite everywhere. Hence, we will first analyze the entropy production treating the Wigner function as a probability distribution over phase space, and using the classical analysis to understand entropy production. The equilibrium distribution is given by 
\begin{equation}
    W_{\rm eq} \propto \exp \left(-\frac{\gamma}{g}x_c^2 - \frac{2\gamma^2}{gk} x_c p_q - \frac{\gamma}{k g}\left(\frac{1}{m} + \frac{\gamma^2}{k} \right)p_q^2 \right) \ .
\end{equation}
The currents are given by 
\begin{equation}
    J_{x_c} = -\frac{g}{2} \frac{\partial W}{\partial x_c} + \frac{p_q}{m} W \qquad J_{p_q} = -\gamma p_qW - V'(x_c)W \qquad \frac{\partial W}{\partial t} = -\partial_{x_c} J_{x_c} - \partial_{p_q} J_{p_q} \ .
\end{equation}
We want to decompose the current into reversible and irreversible pieces \cite{Spinney_2012}: 
\begin{equation}
    \vec{J} = \vec{J}_{\rm rev} + \vec{J}_{\rm irr} \ .
\end{equation}
This is so because the reversible currents do not contribute to entropy production. The irreversible piece, on the other hand, can contribute, and is of our interest . For our situation, the reversible current is given by $J_{{\rm rev}, x_c} = (p_q/m) W, J_{{\rm rev}, p_q} =-V'(x_c) W$, which is just the reversible Hamiltonian phase space current. Thus, the irreversible piece is given by
\begin{equation}
    J_{{\rm irr}, x_c} = -\frac{g}{2} \frac{dW_{\rm eq}}{dx_c} \qquad J_{{\rm irr}, p_q} = -\gamma p_qW_{\rm eq} \ ,
\end{equation} 
and hence the housekeeping entropy production is given by \cite{Seifert:2012pkl}
\begin{equation}
    \dot{S}_{\rm tot} = \int dx_c\, \int dp_q\, \frac{1}{W_{\rm eq}} \vec{J}_{\rm irr} \cdot D \cdot \vec{J}_{\rm irr} = \int dx_c\, \int dp_q \, \frac{1}{W_{\rm eq}} \frac {J_{{\rm irr}, x_c}^2}{D_{xx}}  + \int dx_c\, \int dp_q \, \frac{1}{W_{\rm eq}} \frac {J_{{\rm irr}, p_q}^2}{D_{pp}}
\end{equation}
where $D$ is the diffusion matrix in the $(x,p)$ space, which for us has the only non-zero entry as $D_{xx} = g/2$. Now, formally, we have that $D_{pp} = 0$, and hence $\dot{S}_{\rm tot} = \infty$. This infinite entropy production \footnote{Formally speaking, the infinity is due to the KL divergence between forward and backward paths blowing up, since $p$ is time irreversible. This leads to $\dot{S}_{\rm hk} = \infty$. See the end of \cref{sec:classical_stoch_with_pot} for further details.} is an artifact of the background geometry picking an absolute direction of time, giving rise to a non-zero Hubble friction. Since entropy production measures time-irreversibility, the time dependence of the background will define an absolute arrow of time.

However, as discussed in \cref{sec:quantum_stoch_with_potential} , the existence of a NESS in a quantum theory does not contradict expectations from the calculation of correlators. A local observer will typically measure a single operator $a = f(x,p)$, whose probability we wish to determine. For example, when one measures only the value of $x$ in the overdamped regime, one can systematically eliminate the effect of the $p$ variable leading to a Markovian random walk with no housekeeping entropy production (this is the regime where $V''(x) \ll m \gamma^2$ as we have analyzed in \cref{sec:noise_in_x}) \cite{Green:2025hmo}, recovering back the well-known stochastic inflation formalism. Furthermore, for the Gaussian theory, since this formalism is exact ($V''(x) \ll m \gamma^2 \implies k \ll m \gamma^2$ which is independent of $x$ and always true once parameters are fixed), we have no visible entropy production for observing just the $x$ path and time reversal symmetry is unbroken. Therefore, the observation of the NESS behavior requires comparing individual paths, something that is hard to operationally observe in quantum mechanics.

At face value, the above result might seem to contradict known properties of the static patch of dS~\cite{Spradlin:2001pw}. Most significantly, $d$+1 dimensional dS can be analytically continued to (and from) the ($d+1$)-sphere. As a result, it enjoys a KMS symmetry related to the periodicity in Euclidean time. Normally, the KMS symmetry (see e.g.~\cite{Bros:1995js,Akhmedov:2020qxd} for discussion) would force detailed balance to arise, in contradiction to the NESS behavior that arises (unambiguously) from the evolution equation for the Wigner evolution~\cite{Nambu:1991vs,Li:2025azq,Green:2025hmo,Calderon-Figueroa:2025dto,Cespedes:2026fdp,Li:2026lwl,Christie:2026dwx}. However, given a scalar field $\varphi(\x,t)$ in the flat slicing, the operators that define the Wigner position and momentum variables measured by a worldline are given by~\cite{Green:2025hmo}
\beq
\begin{aligned}
\varphi(\vec{x}=0) & =\bar{\varphi}=\int \frac{d^3 k}{(2 \pi)^3} \varphi(\vec{k},t) \\
\pi(\vec{x}=0) & =\bar\pi=\frac{1}{N_k} \int \frac{d^3 k}{(2 \pi)^3} K\left(\frac{k^2}{\Lambda(t)^2}\right) \pi(\vec{k},t) \ ,
\end{aligned}
\eeq
where $\varphi(\k,t)$ and $\pi(\k,t)$ are the fourier transforms of the field and its conjugate momentum, $K(x)$ is our regulator\footnote{These expressions are derived in~\cite{Green:2025hmo} from exact RG, where the regulator $K$ appears in the action. The field $\bar\varphi$ does not explicitly contain a regulator $K$ because its fluctuations are already suppressed in the path integral.} with time-dependent cutoff $\Lambda = a(t) H$, and $N_k \propto a^3(t)$ the number of modes that have crossed the horizon, starting from some initial time $t_0$. These operators show the time translation symmetry along the worldline is explicitly broken by the initial time $t_0$ and the cutoff $\Lambda=a(t)H$.

The fact that a light scalar field does not respect the symmetries of dS has been known since the early studies of QFT in curved space~\cite{Allen:1985ux}. However, in the presence of a potential $V(\phi)$, the correlators of the field $\bar \varphi$ alone have long been known to recover detailed balance and de Sitter invariance in equilibrium~\cite{Starobinsky:1994bd}. This is a consequence of stochastic inflation after eliminating the momentum variable. Equilibrium eliminates the memory of $t_0$ and any explicit time dependence for the equal-time correlators. The same is not true of the momentum variable - it does not respect the KMS symmetry in the free theory or in the stochastic description (as we have seen).

The infinite entropy might similarly cause concern, given that the de Sitter entropy is a fixed constant value $S_{\rm dS}$~\cite{Gibbons:1977mu}. Of course, our calculation is, strictly speaking, only applicable in fixed dS where $S_{dS} = \infty$. Moreover, the explicit worldline construction of~\cite{Chandrasekaran:2022cip} similarly shows an infinite additive constant to the entropy in the static patch. Our growing entropy can be understood as splitting this infinite constant into a finite time-dependent piece from the IR modes and an infinite UV piece that is not visible to these KMS symmetry breaking operators. At finite $\Mpl$, as in~\cite{Chandrasekaran:2022cip}, there is also no contradiction as measuring $S> S_{dS}$ is not possible for a single local observer due to back-reaction~\cite{Bousso:1999xy}.
\vskip 5pt
{\it Note:} As this work was being completed, several papers appeared on the topic of out-of-time order correlators in dS~\cite{Milekhin:2026tbi,Cui:2026bcd,Chen:2026boh,Harlow:2026pwe}. Some of their results also suggest a breakdown of KMS. While we will leave a comparison of these two observations to future work, we note these results are consistent with our observation that KMS violations are not visible in equal time statistics.

\section{Conclusions}\label{sec:conclusions}

In many complex systems, from statistical mechanics to cosmology, quantum mechanics is formally irrelevant (in the renormalization group sense). Although quantum mechanics plays an essential role in explaining the microscopic contents of the Universe, it's role in macroscopic phenomena is often indistinguishable from classical statistics. The quantum central limit theorem illustrates this tendency, driving systems of identical particles to the same behavior as a classical normal distribution~\cite{Cushen_Hudson_1971}. 

Rare fluctuations are a unique observable that runs counter to conventional renormalization group intuition. Although typical behavior of random systems is driven towards universal distributions, the tails retain knowledge of the microphysics. Understanding these rare events, and their observational signature, provides an IR window into UV physics. As such, we are compelled to understand how microphysics is manifest in these rare fluctuations at a deeper level.

In this paper, we explored how large deviation theory is manifest in quantum walks. First, we demonstrated the interplay between quantum and classical sources of randomness in calculable random walk models. We showed how the classical large deviation instantons and semiclassical WKB configurations are related and showed that there is a smooth cross-over in the behavior of the probability of large deviations. We demonstrated that this behavior is rooted in the minimization the measurement-induced relative entropy of the walk, providing a thermodynamic description of the quantum large deviation results.

Cosmology provides a particularly concrete context where these questions are of fundamental importance. The tail of the distribution is potentially observable in the context of primordial black holes or direct statistical tests. The validity of the WKB approximation or stochastic inflation in predicting these phenomena can be directly relevant for observable predictions. Moreover, the thermodynamics of these fluctuations are tied to the de Sitter entropy. 

Beyond generating non-perturbative statistical predictions, one of the goals of this work is to better understand the relationship between semi-classical gravity, stochastic inflation, and perturbative quantum gravity. These fields have undergone significant progress over the past decade, and yet remain largely disconnected. In the examples studied in this paper, large deviations puts the origin and meaning of semi-classical physics on a solid foundation and points towards the connections to thermodynamics. It is natural to hope that this pattern will continue as one moves from quantum walks to quantum gravity.

\paragraph{Acknowledgments}
We are grateful to JJ Carrasco, Chang-Han Chen, Prish Chakraborty, Tim Cohen, Yiwen Huang, Tom Hartman, Austin Joyce, Yue-Zhou Li, Geoff Penington, Eva Silverstein, and Qiya Zhang.
We are supported by the US~Department of Energy under grant~\mbox{DE-SC0009919}. 

\newpage
\appendix

\section{Relation Between Classical and Quantum Walks}\label{app:iid}

Throughout the paper, we rely on the intuition that our i.i.d.~quantum walks are mathematically equivalent to classical i.i.d.~random walks when measured in a single basis. In this appendix, we will make this connection as transparent as possible for reference throughout the main text.

First, consider a classical random walk $x = \sum_{i=1}^N s_i$ where $s_i$ are i.i.d.~random variable drawn from a distribution $\nu(s)$. Suppose we wish to calculate the probability of finding ourselves at position $x$ after $N$ stepts, $P(x|N)$. This is just the product of the probability of all the steps, integrating over paths that all leave $x$ as the final position, 
\beq
P(x|N)= \int ds_1 .. \int ds_N \nu(s_1) .. \nu(s_N) \delta \left( x-\sum_{i=1}^N s_i \right) \ .
\eeq
The key to simplify this expression is to write the $\delta$-function via for fourier transform
\beq
\delta \left( x-\sum_{i=1}^N s_i \right) \  = \int \frac{dk}{2\pi} e^{i k (x-\sum_i s_i)} \ .
\eeq
Inserting this into our expression we get
\beq
P(x|N) = \int \frac{dk}{2\pi} e^{i k x} \int ds_1 \nu(s_1) e^{-i k s_1} ..\int ds_N \nu(s_N) e^{-i k s_N}
\eeq
Observing that the fourier transform of $\nu(s)$, $\tilde \nu(k)$, is given by
\beq
\tilde \nu(k) = \int ds \nu(s) e^{-i k s} \ ,
\eeq
we see that the probability distribution is given by 
\beq
P(x|N) = \int\frac{dk}{2\pi} e^{i k x} \tilde \nu(k)^N \ .
\eeq
From here one can also prove the central limit theorem by using the connection between $\tilde \nu(k)$ and the moment generating function for $\nu(s)$.

Now let's see the comparison with our canonical quantum walk, \cref{eq:q_model}, given here as
\beq
\rho(x,x') = \bigg\langle x \bigg| {\rm Tr}_i \exp \left(i \sum_i \hat p \hat s_i \right) \rho_0 \otimes \chi^{\otimes N}  \exp \left(-i \sum_i \hat p \hat s_i \right) \bigg| x' \bigg\rangle \ ,
\eeq
Just like the classical walk, it is much easier to use momentum space to write this expression in detail, namely
\beq
\rho(x,x') = \int \frac{dp dp'}{(2\pi)^2} e^{-i p x+ i p' x'} \rho_0(p,p') \left( {\rm Tr} e^{ip s} \chi(s,s') e^{-i p' s'} \right)^N\ \ .
\eeq
Now evaluating the trace using $s=s'$, we find the result
\beq
\rho(x,x') = \int \frac{dp dp'}{(2\pi)^2} e^{-i p x+ i p' x'} \rho_0(p,p') \left(\int ds e^{i(p-p')s} \chi(s,s) \right)^N\ .
\eeq
Now suppose that the initial density matrix is diagonal in the position basis so that 
\beq
\rho_0(x,x') \propto \delta(x-x')q(x) \to \rho_0(p,p') = \tilde q(k=p-p') 
\eeq
where we defined $k = p-p'$ and $\tilde q(k)$ is the fourier transform of $q(x)$. Now we define $x'= x + \Delta x$ and integrate over $k' = p+p'$ to see that
\beq
\rho(x,x') = \delta(\Delta x) \int \frac{dk}{(2\pi)} e^{ik x} \tilde q(k) \left(\int ds e^{ik s} \chi(s,s) \right)^N\ .
\eeq
Defining $\chi(s,s) =\nu(s)$ we can see this is 
\beq
\rho(x,x') = \delta(\Delta x) \int \frac{dk}{(2\pi)} e^{ik x} \tilde q(k) \tilde \nu(k)^N
\eeq
Finally, if we demand that the start at $x=0$ for $N=1$, then $\tilde q(k) = 1$ and we recover the classical i.i.d. random walk
\beq
\rho(x,x') = \delta(\Delta x) \int \frac{dk}{(2\pi)} e^{ik x} \tilde \nu(k)^N \ ,
\eeq
up to the $\delta$-function that enforces the density matrix is diagonal.

\section{Quantum Relative Entropy and Measurement}\label{app:entropy}

Naively, one might have expected the quantum relative entropy to have appeared in the derivation of rate function for a quantum walk in place of the classical relative entropy from Sanov's theorem. In this appendix we will highlight the specific reasons that the min-relative entropy appears, rather than the quantum relative entropy.

\subsection{Pure State Problems}

Say we have a collection of $N$ identical spins, all pointing in the $x$ direction, $\ket{\psi} = \ket{+}^{\otimes N}$. Now, we want to measure the distributions of the outcomes when measuring in the $z$ basis. The problem should be similar to the classical coin tossing experiment - each coin flip is equivalent to measuring the spin in the $z$ basis. Now, classically, we know that the probability of an empirical distribution will be given by Sanov's theorem
\begin{equation}
    P(\mu(x)) = \exp(-N D_{\rm KL}(\mu(x)\|p(x))) \ ,
\end{equation}
where $p(x) = 1/2 [\delta(x - 1) + \delta(x+1)]$. For the quantum theory, we want to generalize this to describe density matrices. A natural guess would be to say that the probability of measuring an empirical density matrix $\rho$ is given by 
\begin{equation}
    P(\rho) = \exp(-N D_{\rm KL}(\rho\|\sigma)) \ ,
\end{equation}
where $\sigma = \ket{+}\bra{+}$ is the true distribution, and here the relative entropy is defined as $D_{\rm KL}(\rho|\sigma) = \Tr[ \rho \log \rho - \rho \log \sigma]$. However, for the case of $N$ spins, if we took this formula at face value, then since $\ket{+}$ is a pure state, we would get that 
\begin{equation}
    D_{\rm KL}(\rho\|\sigma) = \infty \iff \rho \neq \sigma \ .
\end{equation}
The explanation for this is simple. Since the relative entropy is basis agnostic, when designing the experiment we should not measure in the $z$ basis. Rather, we should measure in the $x$ basis. There, we are guaranteed to get $+1$ for every measurement, and hence we should only get one answer $p(x) = \delta(x - N)$ at the end of the measurement. The key idea here was that we never really enforced our limitation of being able to measure only in the $z$ basis. 

Thus, while there are generalizations of this quantum Sanov's theorem that are used in quantum metrology, this is insufficient for us. Rather, we want to be able to describe what the probability of various distributions is, when we do not have access to the full density matrix. As we saw in \cref{sec:quantum_sanov}, the measurement-induced relative entropy is the correct object, which reproduces the classical Sanov's theorem we expect in the case of $N$ spin $1/2$ particles.

\subsection{Distributions versus Single Shot Probability}

The challenge of deriving a single large deviation result for quantum systems is that the form of the answer depends precisely on the measurement we are performing. While aspects of the role of measurement are uniquely quantum, the origin can be traced to our classical model, \cref{eq:class_model}. 

First, suppose we take $q(x_0) = \delta(x_0)$ so that we have a conventional i.i.d.~random walk. In this situation, the empirical distribution for a single walker taking $N$-steps is equivalent to $N$-walkers taking a single step with a distribution $p(x)$. In this sense, we can regard $P(\{n_s \})$ as the probability of finding a histogram for these $N$ walkers. Concretely, suppose the probability of starting at a point $x_0$ and ending at a point $x$ after $L$ steps is $G_L(x|x_0)$, then the probability that $N$ particles starting at $x_0$ will end in a distribution $n_x$ is
\beq
P_{N,S}(\{ n_k \}) =\exp\left( - N \sum_{x\in {\cal X}} \frac{n_x}{N} \log \frac{n_x}{N G_L(x|0)} \right) \ ,
\eeq
where ${\cal X}$ is the set of possible final locations (which is, for $L\neq 1$, is distinct from the set of possible values for a single step, ${\cal S}$).
For the case of a single step, $L=1$, $G_1(x,0) = p_x$ and this is identical to the empirical distribution of steps for a single walker taking $N$ steps (since for $L=1$, ${\cal X} = {\cal S}$).

Now let's allow for $q(x_0) \neq \delta(x_0)$. Suppose we want to know the probability of starting at a distribution $m_\ell \to \mu_0(\x_0)$, where $\ell \in {\cal L}$ is the set of possible initial values, and ending at a distribution $n_x \to \mu(x)$ after $L$ steps. Now the probability is
\beq
P_{N,S}(\{ n_x \}, \{ m_\ell \}) = \exp\left(- N \sum_{\ell \in {\cal L}} \log \frac{m_\ell}{N} \log \frac{m_\ell}{N q_\ell}  - N \sum_{\ell\in {\cal L}, x\in {\cal X}} \frac{m_\ell n_x}{N^2} \log\frac{n_x}{N G_L(x,\ell)} \right) \ .
\eeq
The key feature of this formula is that the first term, which is the probability of the initial distribution $m_\ell$ appears with a factor of $N$, whereas for a single walker taking $N$ steps, \cref{eq:class_prob}, it does not. This is significant because for an $N$ particle distribution, $q(x_0)$ appears like a thermodynamic quantity while for a single walker $q(x_0)$ would typically vanish from the rate function in the thermodynamic limit ($N\to \infty$) .

The purpose of this observation is to further highlight that the relationship between the probability of finding a walker at a point $X$ after $N$ steps is not mathematically equivalent to measurements of the entropy or KL-divergence from an distribution of $N$ particles, only in special circumstances. 

It is for precisely this reason that the quantum random walk is determined by the measurement-induced relative entropy, $D_{\cal M}$ and not the relative entropy $D_{\rm KL}$. The relative entropy is instead what is important for hypothesis testing: imagine we are give $N$ copies of an unknown density matrix $\chi_N = \chi^{\otimes N}$ and we want to know the probability that $\sigma$ is the true density matrix. The best we can do is if we measure $\chi$ in the basis where it is diagonal, so that
\beq
P(\sigma | \chi_N) = \exp\left( N {\rm Tr} \sigma(\log \sigma- \log \chi )  \right) = \exp(-N D_{\rm KL}(\sigma \|\chi) ) \ .
\eeq
As we saw in the case of a single spin, this is particularly relevant for a pure state where $\chi  = |\psi \rangle \langle \psi| $ and $D_{\rm KL}(\sigma |\chi) = \infty$. The intuitive reason is that, for a pure state, we pick any eigenvalue of that state, $\lambda$, with probability 1. Unless $\sigma = \chi$, there is zero probability of producing any of the eigenvalues $\lambda'\neq \lambda$. This result, particularly views asn hypothesis testing, is  refereed to as either quantum Sanov's theorem or quantum Stern's lemma.

In contrast, a quantum walk is sensitive to the ``empirical" density matrix elements in the basis of the walk. We are not hypothesis testing in the same sense and therefore we find a different entropy appears. Because the observer only measures $X$, there is no freedom for the observer to measure matrix elements that would lead to the relative entropy. For example, for a walk in the position basis, there is no way for a measurement of $X$ to tell if the steps are drawn from a the ground state of a harmonic oscillator, or from a classical normal distribution.

The general takeaway for any quantum walk or Markov chain is that the entropy relevant to the walk is determined by $\Sigma$ and not under the control of the observer. The result is that the quantum fidelity and measurement-induced relative entropy always control the probability of finding $X$ in particular locations or states.

\section{Derivation of Open Influence Action for noise in $x$}\label{app:action_derivation}

In this appendix, we give a more rigorous derivation of \eqref{eq:action_noise_in_x}, starting from the underlying Langevin equations. We have the Langevin equations
\begin{equation}
    \dot{x} = \frac{p}{m} + \xi \qquad \dot{p} = - \gamma p- V'(x) \ .
\end{equation}
Here, since the noise is in $x$, the operator that implements the action of the noise on the walker is given by $\exp(i \hat{p} \sum_i s_i)$, where $s_i$ are the step sizes as in \eqref{eq:quantum_walk}. To diagonalize this operator and trace out the environment, it is more convenient to include $x, p$ both in our action. Thus, instead of starting with the Feynman path integral with the momentum variable integrated out, we start with the path integral representation with the momentum still in place. For a closed system, this is given by 
\begin{equation}
    Z = \int {\cal D} p {\cal D} x \exp \left(i \int dt\, (p \dot{x} - H(p, q) \right)  \ .
\end{equation}
Then, the density matrix for the closed system evolves according to 
\begin{equation}
\begin{aligned}
    \rho &= \int {\cal D} p {\cal D} p'{\cal D}x {\cal D}x' \exp \left(i \int dt\, (p \dot{x} - H(p, x))  - i \int dt (p' \dot{x}' - H(p', x') \right) \rho_0 \\
    &= \int {\cal D} p {\cal D} p'{\cal D}x {\cal D}x' \exp(i S_0) \ .
\end{aligned}
\end{equation}
where $S_0$ is the action of the closed system. For the rest of the analysis, it will be convenient to introduce variables $p_c = (p + p')/2, p_q = p - p'$ in addition to $x_c, x_q$. These variables again have the interpretation that $p_c$ acts as the classical momentum, while $p_q$ implements noise/quantum effects. Now, we want to introduce friction into our action. This is given by modifying the action via
\begin{equation}
    S = S_0 - \int \gamma p_c x_q \ ,
\end{equation}
where $p_c = (p + p')/2, x_q = x - x'$. This is the first non-unitary piece which we add. As a check, this gives us back the correct equations of motion
\begin{equation}
    \frac{\delta S}{\delta x_q} = 0 \implies \dot{p}_c = -\gamma p_c - V'(x_c) + {\cal O}(x_q^2) \ .
\end{equation}
Alternatively, one can see that this term appropriately gives us the $\gamma \dot{x}_c x_q$ term as in \eqref{eq:Schwinger_Keldysh_thermal} on integrating out $p_c$. Next, we want to add the effect of the Gaussian noise in $x$. To understand this, we couple the walker explicitly to one particle.  To implement the random walk due to this particle, we have to act with the translation operator, leading to the density matrix of the full system as $\rho_f = \exp(i \hat{p}_w \hat{s}) (\rho_w \otimes \rho_s) \exp(-i \hat{p}_w \hat{s})$, where $w, s$ refer to the walker and the particle degrees of freedom respectively. Tracing over the particle, the reduced density matrix for the walker is given by the Gaussian integral
\begin{equation}
\begin{aligned}
    \rho &= \int d \xi \exp(i (p - p') \xi) \exp(-\xi^2/2g) \rho_w\\
    &= A\exp \left(- \frac{g(p - p')^2}{2} \right) \rho_w \ .
\end{aligned}
\end{equation}
Finally, implementing the noise acting at each time step and reintroducing the Hamiltonian and friction terms, the path integral for our model is given by
\begin{equation}
    \rho =  \int {\cal D} p {\cal D} p'{\cal D}x {\cal D}x' \exp \left(- \int dt\, \frac{g(p - p')^2}{2} \right) \exp(i S) \rho_0 \ .
\end{equation} 
Thus, we see that the noise in $x$ introduces the damping term $\exp(-gp_q^2/2)$ in contrast to the noise in $p$ case which introduced $\exp(-g x_q^2/2)$.  To further simplify, we note that
\begin{equation}
\begin{aligned}
    H(p, x) - H(p', x') &= \frac{p^2 - p'^2}{2m} + (V(x) - V(x')) \\
    &= \frac{1}{m}p_q p_c + V \left(x_c + \tfrac{1}{2} x_q \right) - V\left(x_c - \tfrac{1}{2} x_q \right) \\
    &= \frac{1}{m}p_q p_c + \Delta V\ ,
\end{aligned}
\end{equation}
where we have called $ V(x_c + \frac{1}{2} x_q) - V(x' - \frac{1}{2} x_q) = \Delta V$ for notational simplicity. We also have
\begin{equation}
    p \dot{x} - p' \dot{x}' = \frac{1}{2} [ (p + p)' (\dot{x} - \dot{x}') + (p - p')(\dot{x} + \dot{x}') ] = p_c \dot{x}_q + p_q \dot{x}_c \ .
\end{equation}
Thus, we get
\begin{equation}
\begin{aligned}
    \rho = &\int {\cal D} p_q {\cal D} p_c{\cal D}x_q {\cal D}x_c \exp(- g \int dt\frac{p_q^2}{2} - i\gamma \int  dt \,p_c x_q) \cdot \\
    &\exp \left[i \int dt (p_c \dot{x}_q + p_q \dot{x}_c) - i \int dt \bigg(\frac{1}{m}p_q p_c + \Delta V\bigg) \right] \rho_0 \ .
\end{aligned}
\end{equation}
Now, we can integrate out $p_c, p_q$, leading to
\begin{equation}
    \begin{aligned}
        \rho &= \int {\cal D} p_c {\cal D}x_q {\cal D}x_c \exp \left[ - \frac{1}{2g} \int dt\, \left( \dot{x}_c - \frac{p_c}{m} \right)^2 \right] \exp \left[i \int dt\, (p_c (\dot{x}_q - \gamma x_q) - \Delta V) \right] \rho_0 \\
        &=\int {\cal D}x_q {\cal D}x_c \exp \left[ - \frac{g m^2}{2} \int dt\, ( \dot{x}_q - \gamma x_q )^2\right]  \exp \left[i m\int dt\, (\dot{x}_c (\dot{x}_q - \gamma x_q) - \Delta V) \right] \rho_0\\
        &= \int \mathcal{D}x_c {\cal D}x_q \exp\bigg( i \int dt\, \left[ m \dot{x}_{q} \dot{x}_{c} - m\gamma \dot{x}_{c} x_{q} - \left(V(x_c + \tfrac{1}{2} x_q) - V(x_c - \tfrac{1}{2} x_q ) \right) \right]\\
        & \qquad \qquad - \int dt\, \frac{g m^2\gamma ^{2} }{2}x_{q}^{2} - \int dt\, g m^2\frac{\dot{x}_{q}^{2}}{2}  \bigg) \rho_0 \ .
    \end{aligned}
\end{equation}
This is the required result for \eqref{eq:action_noise_in_x}.

\section{Time Reversibility}
\label{app:TimeReversibility}

Although the microscopic dynamics underlying a thermodynamic process is time-reversal symmetric, macroscopic irreversibility emerges when we depart from equilibrium. Crooks' fluctuation theorem makes this comparison exact at the level of individual trajectories \cite{Crooks98}. Extending the same consideration to nonequilibrium steady states, we arrive at the concept of housekeeping entropy, which isolates the time asymmetry sustained by persistent probability currents even when the state itself is stationary \cite{Oono_Paniconi,Hatano2001}. We review these ideas below, as they furnish the operational measure of time asymmetry used throughout this work.

\subsection{Crooks' Fluctuation Theorem}
\label{app:Crooks}

Consider a system with configuration $x$, driven by an external protocol $\lambda(t)$ over $t \in [0, \tau]$, with Hamiltonian $H(x, \lambda)$. The system starts in equilibrium at $\lambda(0)$,
\begin{equation}
    \pi(x; \lambda) = \frac{e^{-\beta H(x, \lambda)}}{Z(\lambda)} , \quad
    F(\lambda) = -\beta \log Z .
    \label{eq:EquilibriumDist}
\end{equation}
We define the forward process with the protocol $\lambda(t)$, where $t$ is taken from $0 \to \tau$ starting at $\pi(\cdot; \lambda(0))$. The reverse process is affected by the time-reversed protocol $\tilde{\lambda}(t) = \lambda(\tau - t)$, initialized from $\pi(\cdot; \lambda(\tau))$. For instance, in the textbook example of a gas compressed by a piston, the position of the piston would be $\lambda(t)$. We will discretize time, with $i$ indexing the time slices spaced $\Delta t$ apart, so $t = i \Delta t$ and $\tau = n \Delta t$. Then, a particular path $\gamma \equiv ( x_0, \cdots, x_n )$ through phase space is realized as
\begin{equation}
    x_0 \overset{\lambda_1}{\longrightarrow} x_1 \overset{\lambda_2}{\longrightarrow} x_2 \overset{\lambda_3}{\longrightarrow} \cdots \overset{\lambda_n}{\longrightarrow} x_n .
    \label{eq:ForwardPath}
\end{equation}
We only have control over $\lambda_i$, and $\gamma$ can be any path that arises as we apply this protocol. At time $t=0$ the system is in state $x_0$ and the control parameter is $\lambda_0$. The time evolution of the system over the interval $\Delta t$ consists of two substeps. First, the control parameter is moved to a new value $\lambda_1$. This takes an amount of \textit{work} $H(x_0, \lambda_1) - H(x_0, \lambda_0)$. Then the state of the system evolves, at constant $\lambda_1$, to the next state in in the path, $x_1$. During this evolution the system exchanges a quantity $H(x_1, \lambda_1) - H(x_0, \lambda_1)$ of \textit{heat} with the reservoir. These steps are repeated over the entire path.

The total work performed on the system $W$, the total heat exchanged with the reservoir, $Q$,\footnote{We use the convention that heat leaving the system is positive. If $H(x_{i+1}, \lambda_{i+1}) < H(x_i, \lambda_{i+1})$ the system relaxes to a lower energy state, dumping $Q_i$ amount of energy into the reservoir.} and the total change in energy, $\Delta E$, are given by
\begin{subequations}
\begin{align}
    W &= \sum_{i=0}^{n-1} H(x_i, \lambda_{i+1}) - H(x_i, \lambda_i) , \\
    Q &= \sum_{i=0}^{n-1} H(x_i, \lambda_{i+1}) - H(x_{i+1}, \lambda_{i+1}) , \\
    \Delta E &= W - Q = H(x_n, \lambda_n) - H(x_0, \lambda_0) .
\end{align}
\label{eq:Energetics}
\end{subequations}
Next, we assume that the evolution of the system is Markovian. Then the probability of making a transition between two states depends only on the state of the system at time $t$, and not on the previous history of the system. Thus, the probability of the forward path from \cref{eq:ForwardPath} factorizes into
\begin{equation}
    P_{F}[\gamma] = \pi(x_0; \lambda_0) \prod_{i=0}^{n-1} p(x_{i+1} | x_i ; \lambda_{i+1}) .
\end{equation}
Similarly the conjugate reverse trajectory $\gamma^\dagger$ under the reverse protocol $\tilde{\lambda}(t)$,
\begin{equation}
    x_0 \overset{\lambda_1}{\longleftarrow} x_1 \overset{\lambda_2}{\longleftarrow} \cdots \overset{\lambda_{n-1}}{\longleftarrow} x_{n-1} \overset{\lambda_n}{\longleftarrow} x_n ,
    \label{eq:ReversePath}
\end{equation}
has a probability,
\begin{equation}
    P_{R}[\gamma^\dagger] = \pi(x_n; \lambda_n ) \prod_{i=0}^{n-1} p(x_i | x_{i+1} ; \lambda_{i+1}) .
\end{equation}
The key microscopic ingredient is that the dynamics at any fixed $\lambda$ satisfies detailed balance with respect to $\pi(\cdot; \lambda)$:
\begin{equation}
    p(x_{i+1} | x_{i} ; \lambda_i) \pi(x_{i} ; \lambda_i) = p(x_{i} | x_{i+1} ; \lambda_i) \pi(x_{i+1} ; \lambda_i) ,
    \label{eq:DetailedBalance}
\end{equation}

Detailed balance allows us to replace ratios of the forward and reverse kernels with ratios of equilibrium densities. In turn, this allows the ratio of path probabilities to be expressed in terms of the quantities in \cref{eq:Energetics}. Then,
\begin{align}
    \frac{P_{F}[\gamma]}{P_{R}[\gamma^\dagger]}
        &=
        \frac{\pi(x_0; \lambda_0)}{\pi(x_n; \lambda_n )}
        \prod_{i=0}^{n-1} \frac{p(x_{i+1} | x_i ; \lambda_{i+1})}{p(x_i | x_{i+1} ; \lambda_{i+1})} \nn \\
        &=
        \frac{\pi(x_0; \lambda_0)}{\pi(x_n; \lambda_n )}
        \prod_{i=0}^{n-1} \frac{\pi(x_{i+1}; \lambda_{i+1})}{\pi(x_{i}; \lambda_{i+1})} \label{eq:RatioOfEqDist} \\
        &=
        \frac{e^{\beta F_0 - \beta H(x_0, \lambda_0)}}{e^{\beta F_N - \beta H(x_n, \lambda_n)}}
        \exp \left[ \beta \sum_{i=0}^{n-1} \left( H(x_{i} , \lambda_{i+1}) - H(x_{i+1} , \lambda_{i+1}) \right) \right] \nn \\[0.5em]
        &=
        e^{ \beta (-\Delta F + \Delta E + Q) }
        =
        e^{\beta(W - \Delta F)} .
        \label{eq:CrooksFT}
\end{align}
Notice that $W \equiv W[\gamma]$ and $Q \equiv Q[\gamma]$ depend on the specific path taken. The heat generated at each step is a result of the system re-equilibrating to the new value of $\lambda$. As the system relaxes, it loses memory of the previous state from which it started. This is why it is difficult for the system to retrace its evolution when the protocol is reversed.

Recall that the path $\gamma$ is one of many paths the system can take when we apply the protocol $\lambda_i$. That means, at $i=n$, the system is in a state $x_n$ that is is highly probable under $\pi(\cdot; \lambda_n)$. For the system to explore the reverse trajectory from this point it must transition to some $x_{n-1}$ which is a very probable state under $\pi(\cdot; \lambda_{n-1})$, but not so much when then the system is still held at $\lambda_n$ (cf.\ \cref{eq:ReversePath}). Therefore, for a `typical' $\gamma$, we can expect $P_{R}[\gamma^{\dagger}]$ to be much smaller than $P_{F}[\gamma]$. In fact, with $\beta=1$,
\begin{equation}
    S_{\rm tot} \equiv \langle W \rangle - \Delta F
        = \int_\gamma P_{F}[\gamma] \log \frac{P_{F}[\gamma]}{P_{R}[\gamma^\dagger]} \\
        \equiv D_{\rm KL} \left( P_{F}[\gamma] \| P_{R}[\gamma^\dagger] \right)
        \geq 0 . \label{eq:SecondLaw}
\end{equation}
So typical paths, which dominate the average $\langle W \rangle$, usually require more than $\Delta F$ amount of work. This is the Second Law, $\langle W \rangle \geq \Delta F$.

\subsection{Housekeeping Entropy}
\label{app:HousekeepigEntropy}

We now extend the above discussion to the case where detailed balance no longer holds \cite{Hatano2001}. Then, there is a net circulation of probability mass in the triangle, even though snapshots of the probability distribution do not change from one instant to the next. This is called a Non-Equilibrium Steady State (NESS). An example is shown in \cref{fig:NESS_triangle}.

Assume that the system can sustain a NESS at fixed $\lambda$, denoted by $\pi_{\rm ss}(x; \lambda)$. We \textit{define} housekeeping entropy at the site $i$ as the exponent that produces a detailed-balance-like relationship in this setting:
\begin{equation}
    \frac{p(x_{i+1} | x_i ; \lambda_{i+1})}{p(x_i | x_{i+1} ; \lambda_{i+1})}
        \equiv \frac{\pi_{\rm ss}(x_{i+1}; \lambda_{i+1})}{\pi_{\rm ss}(x_{i}; \lambda_{i+1})} e^{s_{{\rm hk}, i}} .
    \label{eq:HousekeepingDefn}
\end{equation}
Following the derivation of \cref{eq:CrooksFT}, but replacing the ratio of transition probabilities with the above,
\begin{equation}
    \frac{P_{F}[\gamma]}{P_{R}[\gamma^\dagger]}
        =
        \frac{\pi_{\rm ss}(x_0; \lambda_0)}{\pi_{\rm ss}(x_n; \lambda_n )}
        \prod_{i=0}^{n-1} \frac{\pi_{\rm ss}(x_{i+1}; \lambda_{i+1})}{\pi_{\rm ss}(x_{i}; \lambda_{i+1})} e^{s_{{\rm hk}, i}} .
\end{equation}
Taking the log, we obtain
\begin{equation}
    \log \frac{P_{F}[\gamma]}{P_{R}[\gamma^\dagger]} = s_{\rm hk}[\gamma] + s_{\rm e}[\gamma] , 
\end{equation}
where the housekeeping and excess entropies are defined as
\begin{gather}
    s_{\rm hk}[\gamma] = \sum_{i=0}^{n-1} s_{{\rm hk}, i}
        = \sum_{i=0}^{n-1} \log \frac{p(x_{i+1} | x_i ; \lambda_{i+1}) \pi_{\rm ss}(x_{i}; \lambda_{i+1})}{p(x_i | x_{i+1} ; \lambda_{i+1}) \pi_{\rm ss}(x_{i+1}; \lambda_{i+1})} , \\[0.5em]
    s_{\rm e}[\gamma]
        =
        \log \frac{\pi_{\rm ss}(x_0; \lambda_0)}{\pi_{\rm ss}(x_n; \lambda_n )} +
        \sum_{i=0}^{n-1} \log \frac{\pi_{\rm ss}(x_{i+1}; \lambda_{i+1})}{\pi_{\rm ss}(x_{i}; \lambda_{i+1})}
        =
        \sum_{i=0}^{n-1} \left[ \log \pi_{\rm ss}(x_{i}; \lambda_{i}) - \log \pi_{\rm ss}(x_{i}; \lambda_{i+1}) \right] . \label{eq:ExcessEntropyWithProtocol}
\end{gather}
We can do two consistency checks: (1) if detailed balance is restored $s_{\rm hk}$ vanishes, so we can set $\pi_{\rm ss} \to \pi = e^{-\beta H + \beta F}$ to recover \cref{eq:CrooksFT}, and (2) if $\lambda$ is held fixed the excess entropy becomes zero and all entropy is from housekeeping---the circulating current implies that trajectories in one direction are much more probable than their time-reversed counterparts, even when the overall density is stationary.

\paragraph{Removing the protocol}
We turn our attention now to a system where $\lambda$ is held fixed. Let the system have a NESS, and assume that we start from a distribution $P_0$ and relax to $P_n$ over $n$ steps, where both of these distributions are different from $\pi_{\rm ss}$. Then,
\begin{align}
    \frac{P_{F}[\gamma]}{P_{R}[\gamma^\dagger]}
        &= \frac{P_0(x_0)}{P_n(x_n)} \prod_{i=0}^{n-1} \frac{p(x_{i+1} | x_i )}{p(x_i | x_{i+1})} \nn \\
        &= \frac{P_0(x_0)}{P_n(x_n)} \frac{\pi_{\rm ss}(x_n)}{\pi_{\rm ss}(x_0)} \prod_{i=0}^{n-1} \frac{p(x_{i+1} | x_i ) \pi_{\rm ss}(x_{i})}{p(x_i | x_{i+1}) \pi_{\rm ss}(x_{i+1})} \nn \\
        &= \frac{P_0(x_0)}{P_n(x_n)} \frac{\pi_{\rm ss}(x_n)}{\pi_{\rm ss}(x_0)} e^{s_{\rm hk}[\gamma]} .
\end{align}
Taking the logarithm, we obtain
\begin{equation}
    \log \frac{P_{F}[\gamma]}{P_{R}[\gamma^\dagger]}
        = s_{\rm hk}[\gamma] +
            \underbrace{
                \log \frac{P_0(x_0)}{\pi_{\rm ss}(x_0)} - \log \frac{P_n(x_n)}{\pi_{\rm ss}(x_n)}
            }_{s_{\rm e}[\gamma]} .
    \label{eq:LogOfPathRatio}
\end{equation}
Notice that the excess entropy is a purely state-dependent quantity now, unlike \cref{eq:ExcessEntropyWithProtocol} where it was dependent on the path. Averaging over all paths,
\begin{equation}
    S_{\rm tot} =
    \int \d x_0 \cdots \d x_n P_{F}[\gamma] \log \frac{P_{F}[\gamma]}{P_{R}[\gamma^\dagger]}
        = S_{\rm hk} + D_{\rm KL} (P_0 \| \pi_{\rm ss}) - D_{\rm KL} (P_n \| \pi_{\rm ss}) ,
\end{equation}
where $S_{\rm hk} \equiv \langle s_{\rm hk} \rangle$ is the average housekeeping entropy over all paths.

\clearpage
\phantomsection
\addcontentsline{toc}{section}{References}
\small
\bibliographystyle{utphys}
\bibliography{Refs}

\end{document}